\documentclass{arxiv}

\begin{document}

\title{Gradient-Free Aeroacoustic Shape Optimization Using Large Eddy Simulation}
\author{Mohsen Hamedi and Brian C. Vermeire\\\\
\textit{Department of Mechanical, Industrial, and Aerospace Engineering}\\
\textit{Concordia University} \\
\textit{Montr\'eal, QC, Canada}}
\date{}

\maketitle

\begin{abstract}

We present an aeroacoustic shape optimization framework that relies on high-order Flux Reconstruction (FR), the gradient-free Mesh Adaptive Direct Search (MADS) optimization algorithm, and Large Eddy Simulation (LES). Our parallel implementation ensures consistent runtime for each optimization iteration, regardless of the number of design parameters, provided sufficient resources are available. The objective is to minimize the Overall Sound Pressure Level (OASPL) at a near-field observer by computing it directly from the flow field. We evaluate this framework across three problems. First, an open deep cavity is considered at a free-stream Mach number of $M_\infty=0.15$ and Reynolds number of $Re=1500$, reducing the OASPL by $12.9~dB$. Next, we considered tandem cylinders at $Re=1000$ and $M_\infty=0.2$, achieving over $11~dB$ noise reduction by optimizing cylinder spacing and diameter ratio. Lastly, a baseline NACA0012 airfoil at $Re=23000$ and $M_\infty=0.2$ is optimized to generate a new 4-digit NACA airfoil at an appropriate angle of attack to minimize the OASPL while ensuring the baseline time-averaged lift coefficient is maintained and prevent any increase in the baseline time-averaged drag coefficient. The OASPL and mean drag coefficient are reduced by $5.7~dB$ and more than $7\%$, respectively. These results highlight the feasibility and effectiveness of our aeroacoustic shape optimization framework.


\end{abstract}

\section{Introduction}

Aeroacoustic optimization has received significant attention in recent years due to its various applications, such as reducing wind turbine noise for widespread deployment, minimizing aviation noise to enhance the comfort of communities near airports, and designing quiet air taxis for urban air mobility, among others. An aeroacoustic shape optimization framework comprises three distinct components. Initially, a flow solver is utilized to capture aerodynamic flow characteristics. Subsequently, an acoustic solver computes noise at the observer(s) based on the acquired aerodynamic flow data, which is omitted in the direct acoustic approach, wherein noise is directly computed within the flow solver. The final component is the optimization algorithm, responsible for identifying candidate designs for each optimization iteration. Various aeroacoustic optimization frameworks are constructed by employing different methods for each of these components.

XFOIL simulations have found extensive application in aeroacoustic shape optimization for aerodynamic analysis \cite{kou2023aeroacoustic, volkmer2018aeroacoustic, jones2000aerodynamic}. While optimization frameworks employing panel methods offer cost-effective exploration of design spaces, panel methods may lack the precision needed for reliably obtaining optimal designs \cite{volkmer2018aeroacoustic}. Thus, more reliable methods should be considered to find optimal designs. An alternative to panel methods is Reynolds-Averaged Navier-Stokes (RANS) simulations, which have previously been used for aeroacoustic shape optimization \cite{klimczyk2023rans, monfaredi2021unsteady, monfaredi2021unsteady2, ricks2020cfd, yuepeng2020aerodynamic}. However, due to the inherent unsteady nature of noise phenomena, the RANS approach can add unwanted dissipation of broadband noise \cite{slotnick2014cfd}. Consequently, scale-resolving techniques, i.e., Large Eddy Simulation (LES), Implicit LES (ILES), and Direct Numerical Simulation (DNS) are of interest. They offer an unsteady and detailed representation of the flow physics and resulting acoustic waves, and are appealing alternatives, albeit with added computational cost \cite{colonius2004computational, marsden2007trailing, marsden2001shape}. The majority of Computational Fluid Dynamics (CFD) codes for simulating unsteady compressible flow, such as OpenFOAM \cite{openfoam}, SU2 \cite{su2, palacios2013stanford}, and CHARLES \cite{charles}, rely on Finite Volume (FV) methods with second-order spatial accuracy. While these methods can handle complex geometries on unstructured meshes and scale to approximately one million cores \cite{million}, they are constrained by a low FLOPS-to-bytes ratio and high indirect memory access, preventing them from fully harnessing the computational power of modern hardware platforms \cite{vincent2016towards}. As specified by CFD 2030 Vision study by the National Aeronautics and Space Administration (NASA) \cite{slotnick2014cfd}, the rapid advancement of HPC has outpaced the capabilities of conventional CFD algorithms, highlighting the need for more advanced approaches aligned with modern computing architectures. The industry-standard FV methods only achieve $3\%$ of the theoretical peak performance on modern hardware architectures \cite{langguth2013gpu}; however, the FR approach \cite{huynh2007flux} is capable of achieving over $55\%$ \cite{vincent2016towards}. In addition, the FR approach has been shown to be suitable for scale-resolving simulations, leveraging the behaviour of its numerical error for ILES \cite{vermeire2016implicit}, and via filtering approaches for highly under-resolved problems \cite{hamedi2022optimized}. Thus, FR proves computationally superior to lower-order FV techniques, with reduced numerical dispersion and dissipation errors on a per degree of freedom basis \cite{abgrall2017high, hesthaven2017numerical, wang2013high}. In this study, our High-ORder Unstructured Solver (HORUS) is used, which employs the FR approach for spatial discretization of the governing equations with ILES for sub-grid scale modelling. 

The emergence of adjoint-based optimization methods \cite{pironneau1974optimum, jameson1988aerodynamic}, characterized by computational cost independence of the number of design variables, has enabled the exploration of large-scale practical problems in aerodynamic optimization \cite{lyu2014aerodynamic}. While a substantial body of literature focuses on steady-state problems, the unsteady nature of numerous aerospace problems, such as aeroacoustics, has received less attention in adjoint-based optimization due to the considerable storage requirements for solving unsteady adjoint equations \cite{zhou2015discrete} and their unconditional instability for chaotic systems \cite{karbasian2022sensitivity}. A more robust alternative for aeroacoustic shape optimization using LES is the gradient-free Mesh Adaptive Direct Search (MADS) algorithm \cite{audet2006mesh, abramson2009orthomads}. Unlike optimization methods reliant on gradient information, MADS operates directly on objective function evaluations, making it inherently robust in problems that accurate gradient computation is challenging. This independence is a significant advantage in unsteady simulations, where the objective function's landscape can change rapidly over time, making gradient information less reliable or even inaccurate. MADS directly optimizes the objective function based on its evaluations, allowing it to adapt to these changes more effectively than gradient-based methods. Additionally, the absence of gradient computations reduces sensitivity to initial guesses, promoting more reliable convergence behavior, particularly in transient or highly dynamic environments. Hence, MADS emerges as a compelling choice for optimizing unsteady problems, offering a versatile and robust approach capable of navigating the complexities of time-dependent simulations. The suitability of MADS, coupled with HORUS, has been demonstrated in the works of Karbasian and Vermeire \cite{karbasian2022gradient} and Aubry et al. \cite{aubry2022high} for aerodynamic shape optimization, and by Hamedi and Vermeire \cite{hamedi2024near} for laminar aeroacoustic shape optimization.

In this study, we present an aeroacoustic shape optimization framework based on the FR approach and the gradient-free MADS optimization algorithm for LES. Building upon our prior work \cite{hamedi2024near}, which assessed this framework for two-dimensional problems at low Reynolds numbers, we extend its application to three-dimensions. To the best of our knowledge, no other studies have integrated the gradient-free MADS optimization with a high-order LES solver. One significant limitation of this framework is its runtime since a high-order LES is performed for each objective function evaluation. The runtime of the optimization problem scales with the number of design parameters, requiring a corresponding number of CFD simulations in each optimization iteration. However, we addressed this challenge by implementing the optimization framework in parallel. This enables concurrent evaluation of candidate designs within each optimization iteration, effectively reducing the runtime of each iteration to that of a single CFD simulation and independent of the number of design parameters, provided sufficient computing resources are available. Additionally, each CFD simulation is performed in parallel on state-of-the-art clusters using Graphical Processing Units (GPUs), highlighting the two-layer parallelism of the proposed optimization algorithm. 

This paper is outlined as follows. The methodology is given in Section \ref{sec:Methodology}. Then, shape of a three-dimensional open cavity is optimized to reduce noise in Section \ref{sec:DeepCavity}, followed by three-dimensional tandem cylinders in Section \ref{sec:TandemCylinders}, and, airfoil shape optimization for noise reduction is performed in Section \ref{sec:NACA}. Finally, the conclusions and recommendations for future work are given in Section \ref{sec:Conclusions}.

\section{Methodology}
\label{sec:Methodology}

This section presents an overview of the methodology employed to solve the unsteady Navier-Stokes equations along with the aeroacoustic shape optimization framework.

\subsection{Governing Equations}

The compressible unsteady Navier-Stokes equations can be cast in the following general form
\begin{equation}
\frac{\partial \pmb{u}}{\partial t} + \pmb{\nabla} \cdot \pmb{F} = 0,
\label{equation_conservation_law}
\end{equation}
where $t$ is time and $\pmb{u}$ is a vector of conserved variables
\begin{equation}
\pmb{u} = 
\begin{bmatrix}
\rho \\
\rho u_i \\
\rho E
\end{bmatrix},
\end{equation}
where $\rho$ is density, $\rho u_i$ is a component of the momentum, $u_i$ are velocity components, and $\rho E$ is the total energy. The inviscid and viscous Navier-Stokes fluxes are 
\begin{equation}
\pmb{F}_{inv,j} (\pmb{u}) = 
\begin{bmatrix}
\rho u_j \\
\rho u_i u_j + \delta_{ij} p \\
u_j ( \rho E + p)
\end{bmatrix},
\end{equation}
and
\begin{equation}
\pmb{F}_{vis, j} (\pmb{u}, \nabla \pmb{u}) =
\begin{bmatrix}
0 \\
\tau_{ij} \\
-q_j - u_i \tau_{ij}
\end{bmatrix} ,
\end{equation}
respectively, where $\delta_{ij}$ is the Kronecker delta. The pressure is determined via the ideal gas law as
\begin{equation}
p = (\gamma - 1) \rho \left( E - \frac{1}{2} u_k u_k \right) ,
\end{equation}
where $\gamma=1.4$ is the ratio of the specific heat at constant pressure, $c_p$, to the specific heat at constant volume, $c_v$. The viscous stress tensor is
\begin{equation}
\tau_{ij} = \mu \left( \frac{\partial u_i}{\partial x_j} + \frac{\partial u_j}{\partial x_i} - \frac{2}{3} \frac{\partial u_k}{\partial x_k} \delta_{ij} \right) ,
\end{equation}
and, the heat flux is
\begin{equation}
q_j = - \frac{\mu}{Pr} \frac{\partial}{\partial x_j} \left( E + \frac{p}{\rho} - \frac{1}{2} u_k u_k \right) ,
\end{equation}
where $\mu$ is the dynamic viscosity and $Pr=0.71$ is the Prandtl number. 

\subsection{Optimization Framework}

In this study, we employed the minimal bases construction of the MADS optimization technique, similar to our previous work \cite{hamedi2024near}, for the open deep cavity and tandem cylinders. However, for the NACA0012 problem, we developed a parallel optimization framework, employing the maximal bases construction of the OrthoMADS algorithm \cite{abramson2009orthomads}, to address the runtime challenges inherent in serial implementation. 

\begin{figure}
\centering
\includegraphics[width=\textwidth]{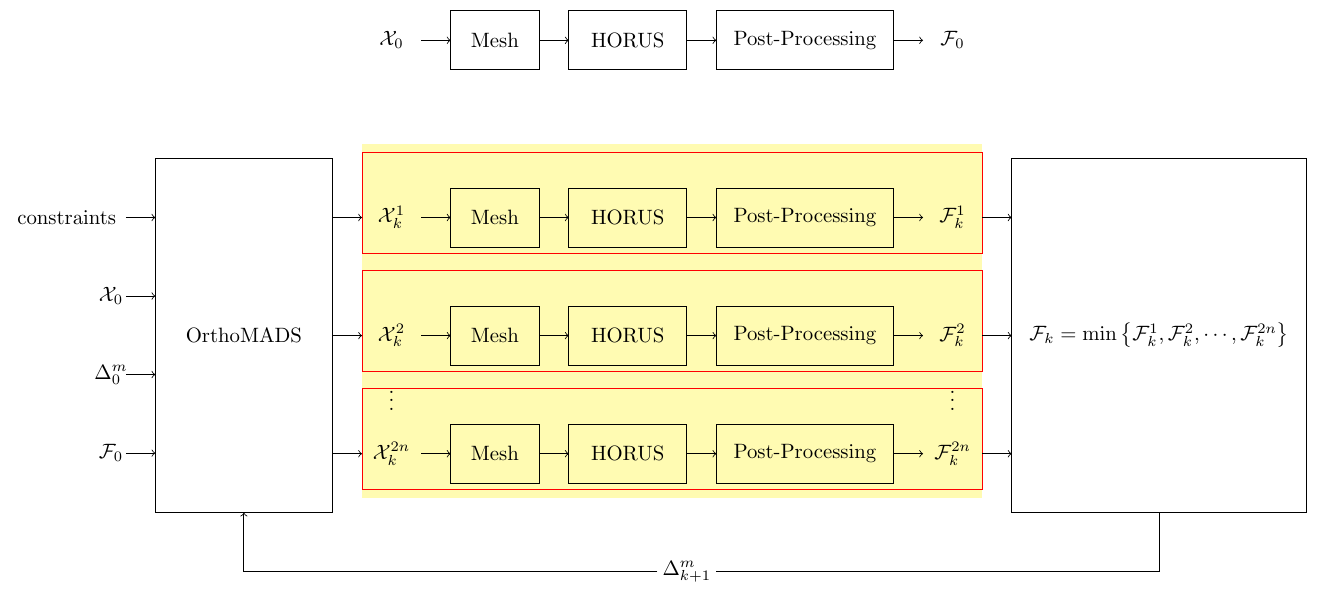}
\caption{Flowchart of the proposed aeroacoustic shape optimization framework.}
\label{fig_mads_flow}
\end{figure}

The flowchart of the proposed aeroacoustic shape optimization is illustrated in Figure \ref{fig_mads_flow}. The process begins by evaluating the baseline objective function, $\mathcal{F}_0$, with the incumbent design set equal to the baseline design, $\mathcal{I}_0 = \mathcal{F}_0$. The optimization algorithm then takes as inputs the problem's constraints, baseline design parameters $\mathcal{X}_0$, initial mesh size parameter $\Delta_0^m$, and the baseline objective function $\mathcal{F}_0$. The mesh size parameter $\Delta^m \in \mathbb{R}_+$ defines the resolution of the design space $\mathcal{D}$, and it guides the selection of design candidates within each optimization iteration along with the poll size parameter $\Delta^p$. The OrthoMADS algorithm employs polling directions orthogonal to each other, generating minimal convex cones of unexplored directions at each iteration, thus enhancing the efficiency of design space exploration \cite{abramson2009orthomads}. Along with Figure \ref{fig_mads_flow}, Algorithm \ref{algorithm_mads} delineates the parallel implementation of the OrthoMADS algorithm.

For the $k$-th optimization iteration, candidate designs are identified, and an automated script generates the mesh for each geometry, with wall surfaces controlled by design parameters that directly influence the geometry. The HORUS is then called to compute the flow field using high-order LES. The objective functions of the candidate designs, $\mathcal{F}_k^i$, are evaluated as a post-processing step and compared to the incumbent design $I_k$. By comparing the objective function of these designs with the incumbent design, both the mesh size parameter and the incumbent design are updated, initiating the next optimization iteration. The optimization process stops when the mesh size parameter falls below $10^{-6}$, and the changes in design parameter values between two consecutive iterations are less than one percent. These criteria indicate the algorithm has successfully converged to an optimal design. 

Notably, the for loop in lines $18-21$ of the Algorithm \ref{algorithm_mads}, which corresponds to the highlighted parts of Figure \ref{fig_mads_flow}, is the most computationally intensive part of the algorithm where a total of $2n$ CFD simulations are conducted, where $n$ is the number of design parameters. Typically, each CFD simulation runs in parallel, and candidate designs are executed sequentially. However, in the proposed parallel implementation of the algorithm, all candidate designs run concurrently, reducing the runtime of $2n$ CFD simulations to that of a single CFD simulation, provided adequate computer resources are available.

\begin{algorithm}[htbp]
	\SetAlgoLined
	\caption{The aeroacoustic shape optimization framework.}
	\label{algorithm_mads}
	$k=0$; \\
	MADS Iteration, $iter=0$; \\
	
	Run Baseline Design; \\
	Evaluate $\mathcal{F}_0$; \\
	Define Incumbent $\mathcal{I}_0 = \mathcal{F}_0$; \\
	
	Define $\Delta^m_0$; \\
	
	\While{True}{
	
		\If{$\Delta^m_k > \Delta^m_0$}{
			$\Delta^m_k = \Delta^m_0$;\\}
	
		\If{minimal positive basis construction}{
			$\Delta^p_k = n \sqrt{\Delta^m_k}$;}
		\If{maximal positive basis construction}{
			$\Delta^p_k = \sqrt{\Delta^m_k}$;}
			
		Identify Candidate Designs, $\pmb{p}^1_k, ..., \pmb{p}^n_k$; \\
		
		\For{$i=1,...,n$}{
			Run HORUS for $\pmb{p}^i_k$; \\
			Evaluate $\mathcal{F}^i_k$; \\}
			
		\eIf{$\min\left\lbrace \mathcal{F}^1_k, ..., \mathcal{F}^n_k \right\rbrace < \mathcal{I}_{iter}$}{
			$\Delta^m_{k+1} = 4 \Delta^m_k$;\\
			$iter$\texttt{+=}$1$;\\
			$\mathcal{I}_{iter} = \min\left\lbrace \mathcal{F}^1_k, ..., \mathcal{F}^n_k \right\rbrace$;\\}
			{$\Delta^m_{k+1} = \frac{1}{4} \Delta^m_k$;\\}
	
		$k$\texttt{+=}$1$; \\
		
		\If{$\Delta^m_k < 10^{-6}$ and $\left| \frac{  \pmb{\mathcal{X}}_k - \pmb{\mathcal{X}}_{k-1} }{\pmb{\mathcal{X}}_{k-1}} \right| < 0.01$}{
		\texttt{break;}}
	}
\end{algorithm}

\subsection{Flow Solver}

The in-house solver, HORUS, is utilized for solving the Navier-Stokes equations, employing the FR approach for spatial discretization. This approach is used to discretize the divergence operator for general advection-diffusion equations of the form shown in Equation \ref{equation_conservation_law}. It is a high-order accurate numerical method first introduced by Huynh \cite{huynh2007flux} in 2007, and extended to multi-dimensions for mixed element types by Wang and Gao \cite{wang2009unifying}. FR is appealing due to its accuracy, generality, robustness, and suitability for modern hardware architectures \cite{vincent2016towards}. Compared to commonly-used low-order numerical methods, FR provides more accurate solutions using fewer degrees of freedom and at reduced computational cost \cite{vermeire2017utility}. We explained the FR approach in more details in our previous work \cite{hamedi2024near}. In this study, the second-order accurate Nasab-Pereira-Vermeire scheme \cite{hedayati2021optimal}, which incorporates an adaptive time-stepping method \cite{vermeire2020optimal}, is used to advance the solution in time. 

\section{Deep Cavity}
\label{sec:DeepCavity}

Flow over an open deep cavity is a classical problem in fluid mechanics and aeroacoustics, and has been the subject of extensive research due to its relevance for a range of engineering applications. Such flows represent simplified versions of the complex dynamics over panel gaps, like those between windows/doors and the fuselage or between control surfaces and wings. These gaps profoundly influence aerodynamics, structural integrity, and noise levels of aircraft. The flow over a cavity is characterized by a complex interplay between the boundary layer, the recirculation zone inside the cavity, and the external flow. The occurrence of self-sustained oscillations of velocity and pressure can induce acoustic noise or strong vibrations. The presence of the cavity can lead to a variety of aerodynamic and aeroacoustic phenomena, such as flow separation, unsteady vortex shedding, and acoustic resonance. Understanding the aerodynamic and aeroacoustic characteristics of flow over a cavity is crucial for optimizing the design and performance of many engineering systems.

\begin{figure}
\centering
\includegraphics[width=0.7\textwidth]{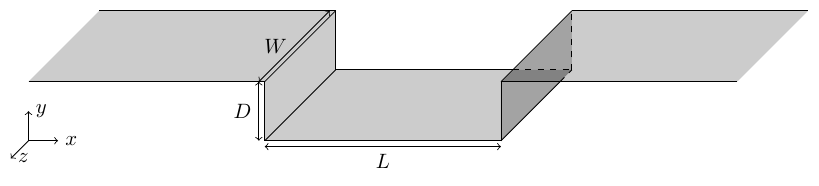}
\caption{The geometry of the three-dimensional open deep cavity.}
\label{fig_cavity_3d_geometry}
\end{figure}

Extensive research has been conducted on two-dimensional cavity flows, leading to favorable agreement between experimental data and numerical two-dimensional simulations. While three-dimensionality is observed in cavity flow experiments, it underscores the significance of conducting three-dimensional cavity flow simulations \cite{maull1963three, rockwell1980observations}. Lawson \cite{lawson2011review} reviewed the experimental and numerical studies of open cavities. Furthermore, the radiated noise from cavity is studied via LES by several researchers \cite{gloerfelt2002noise, gloerfelt2003direct, lai2007three, gloerfelt2004large}. The geometry of a three-dimensional cavity is usually given in terms of length-to-depth, $L/D$, and width-to-depth, $W/D$, ratios, as depicted in Figure \ref{fig_cavity_3d_geometry}. In this section, flow over an open cavity is validated and then the noise at a near-field observer is minimized via the proposed gradient-free shape optimization framework.

\subsection{Validation}  

In this section, we extend our previous work \cite{hamedi2024near} by extruding it in the $z$-direction. The grid convergence study is performed using the time-averaged drag coefficient, and OASPL measured at an observer located $7.16D$ above the cavity's center.

\subsubsection{Computational Details}

\begin{figure}
\centering
\includegraphics[width=0.7\textwidth]{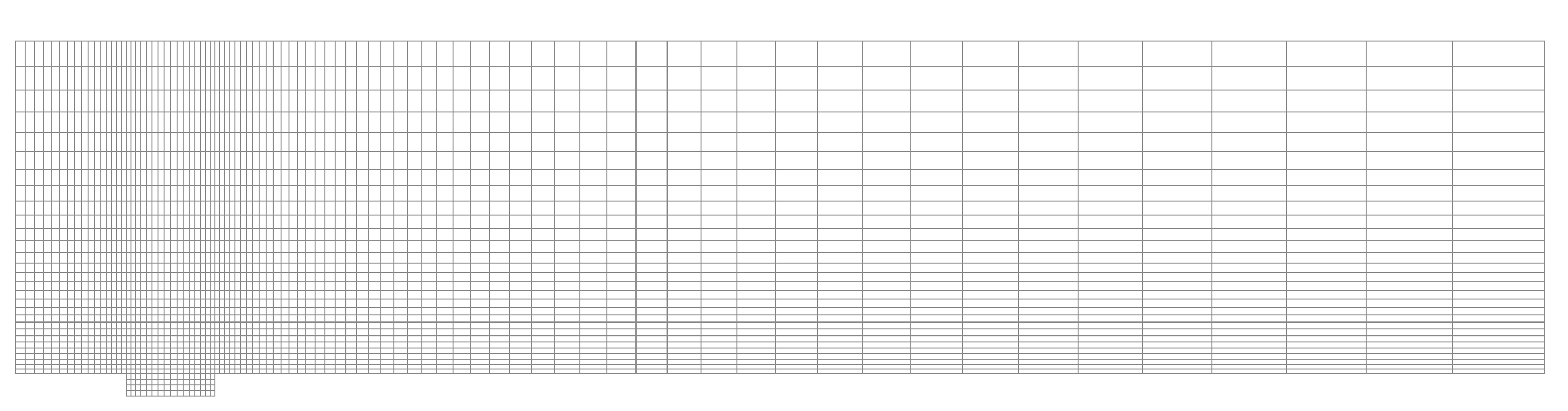}
\caption{The mesh of the three-dimensional open deep cavity.}
\label{fig_cavity_3d_mesh}
\end{figure}

\begin{figure}
\centering
\includegraphics[width=0.7\textwidth]{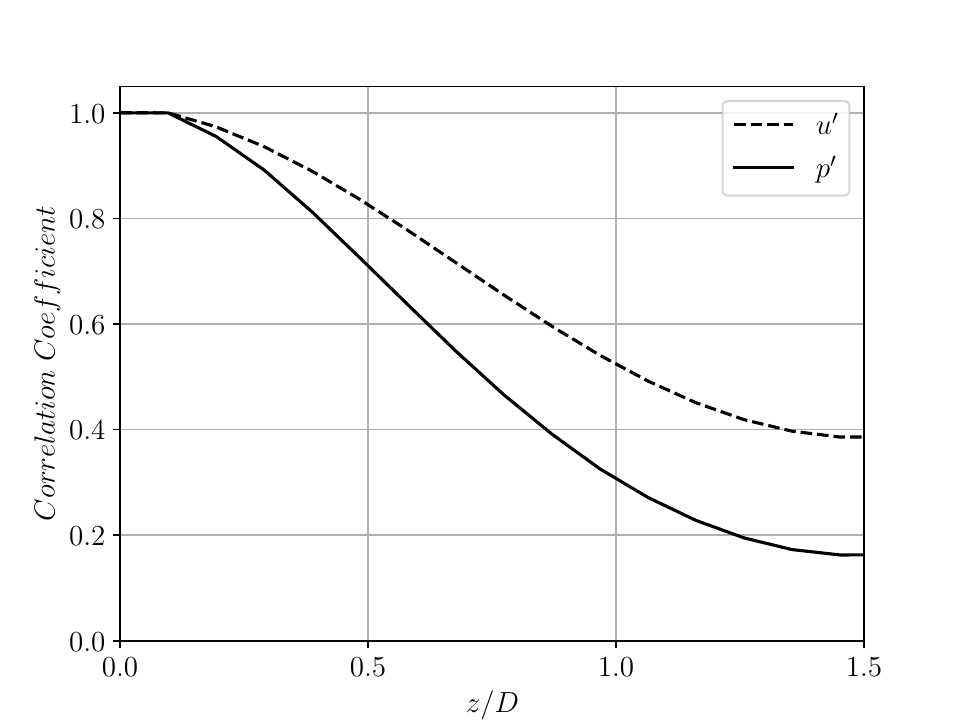}
\caption{The correlation coefficient in the spanwise direction for the three-dimensional open deep cavity.}
\label{fig_cavity_correlation}
\end{figure}

To be consistent with \cite{hamedi2024near}, the open cavity with a length-to-depth ratio of $L/D=4$ is extruded in the $z$-direction with a width-to-depth ratio of $W/D=3$. The Reynolds number, based on the depth of the cavity, is $Re_D=1500$, and the Mach number is $0.15$. To ensure wake mode oscillations, the inlet boundary is placed $5D$ upstream of the cavity inlet, resulting in a boundary layer thickness of $\delta/D \approx 0.2$ at the entrance of the cavity. The outflow boundary is placed $60D$ downstream of the cavity's trailing edge wall, with the last $50D$ acting as a buffer region to eliminate acoustic wave reflections. The computational domain extends to $15D$ in the $y$-direction with the last $5D$ as a buffer region. The grid stretching ratio is $1.05$ and $1.075$ for the resolved and buffer regions, respectively, with a minimum element size of $0.2D$ inside the cavity. A total of $14,652$ hexahedral elements are used. The geometry and mesh of the three-dimensional cavity are shown in Figures \ref{fig_cavity_3d_geometry} and \ref{fig_cavity_3d_mesh}, respectively. The periodic boundary condition is used in the spanwise direction, no-slip boundary conditions are applied at the walls, and Riemann invariant boundary conditions are applied at the inlet and outlet of the computational domain. The simulation is run for $100t_c$, where $t_c=D/U_\infty$, to allow initial transients to disappear and then run for another $400t_c$ to average the statistical quantities. To ensure uncorrelated turbulent fluctuations at a separation of half the domain size, the correlation coefficient of the $x$-component of the velocity perturbation along with that of the pressure perturbation are computed along the spanwise direction and depicted in Figure \ref{fig_cavity_correlation}. The results of the grid independence study are given in the next section.

\subsubsection{Results and Discussion}

The grid independence study is performed by increasing the solution polynomial degree, which increases the resolution of the simulation. The time-averaged drag coefficient and the OASPL at an observer located $7.16D$ above the center of the cavity are computed using solution polynomial degrees of $\mathcal{P}2$, $\mathcal{P}3$, and $\mathcal{P}4$ to show the grid independency.

The drag coefficient is defined as
\begin{equation}
C_D = \frac{F_x}{\frac{1}{2}\rho_\infty U_\infty^2 DW},
\end{equation}
where $F_x$ is the force in the $x$-direction computed on the three cavity walls, $\rho_\infty$ is the free-stream density, and $U_\infty$ is the free-stream velocity. Furthermore, the OASPL is computed using the following equation
\begin{equation}
OASPL = 20 \log \left(\frac{p_{RMS}}{p_{ref}} \right) ,
\label{eq_oaspl}
\end{equation}
where $p_{RMS}$ represents the root-mean-square of the pressure perturbations, defined as
\begin{equation}
p_{RMS} = \sqrt{\frac{\sum_{i=1}^n (p^\prime_i)^2}{n}},
\end{equation}
where $n$ is the total number of time samples, and the pressure perturbation $p^\prime$ is given by
\begin{equation}
p^\prime = p - \bar{p} ,
\end{equation}
with $\bar{p}$ being a vector of cumulative time-averaged pressure signals, each element of which is defined as
\begin{equation}
\bar{p}_k = \frac{\sum_{i=1}^k p_i}{k}, \quad k=1,2,...,n.
\end{equation}
Note that in this formulation, $\bar{p}$, $p^\prime$, and $p_{RMS}$ are vectors, where each element $k$ corresponds to the calculation using $k$ time samples. In Equation \ref{eq_oaspl}, $p_{RMS}$ refers to the last element of the $p_{RMS}$ vector, i.e. $p_{RMS} = p_{{RMS}_n}$.

The time-averaged drag coefficient along with the OASPL at the observer, for different simulations, are given in Table \ref{table_cavity_summary}. $30$ observer points along the span of the cavity are used. The time-averaged pressure and root-mean-squared of the pressure perturbation are computed for each observer point and then spatially averaged to find the OASPL at the observer location.
\begin{table}
\centering
\caption{A summary of grid independence study of the open deep cavity.}
\begin{tabular}{ccc}
\hline
Simulation & $\overline{C_D}$ & OASPL in $dB$ \\
\hline
$\mathcal{P}2$ & $0.1314$ & $112.1$ \\
$\mathcal{P}3$ & $0.1098$ & $113.1$ \\
$\mathcal{P}4$ & $0.1115$ & $113.3$ \\
\hline
\end{tabular}
\label{table_cavity_summary}
\end{table}
These results show that the $\mathcal{P}3$ simulation provides sufficient resolution for this study. 

\subsection{Optimization}

\begin{figure}
\centering
\includegraphics[width=0.7\textwidth]{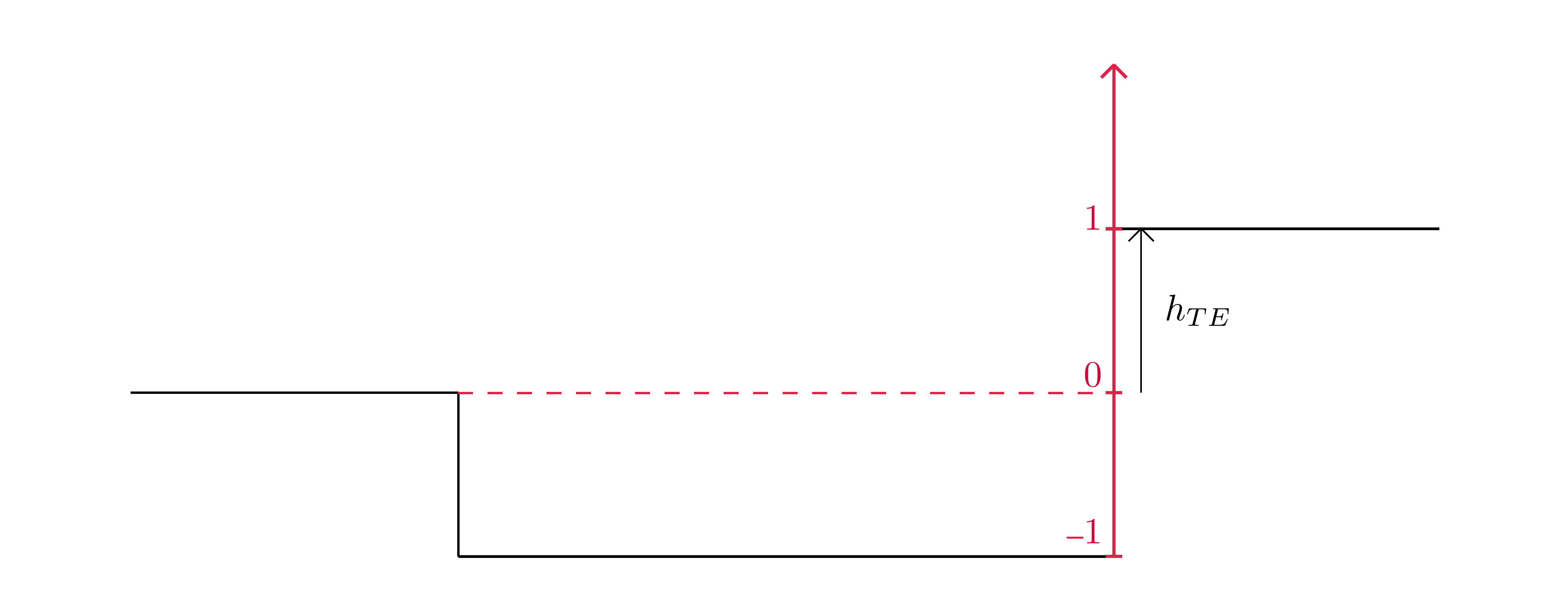}
\caption{The design variable, $h_{TE}$, for the open deep cavity.}
\label{fig_cavity_hte}
\end{figure}

In this section, the noise at the observer point located at $\pmb{x}_{obs} / D = [2, 7.16]$ is minimized by changing the height of the cavity trailing edge wall, $h_{TE}$, depicted in Figure \ref{fig_cavity_hte}. There are other possible shape parameters to minimize the cavity’s noise, such as the length-to-depth ratio. The choice of the design parameter for shape optimization depends on the noise generation mechanism of interest. For instance, if the focus is on exploring the shear layer extending over the cavity without vortex roll-up as the primary mechanism for sound generation, the length-to-depth ratio would be a more suitable design parameter. However, in this study, the focus is on determining whether the vortices become trapped within the cavity or rest on the downstream wall, thereby forming a backward-facing step. Thus, $\pmb{\mathcal{X}}=h_{TE}$ is the design variable and $\pmb{\mathcal{X}}_0 = 0$, while the objective function is $\mathcal{F}=p^\prime_{rms}$. Upper and lower bounds of $-1$ and $4$, respectively, are chosen for the design variable, $h_{TE}$. 

\subsubsection{Results and Discussion}

\begin{figure}
\centering
\includegraphics[width=0.7\textwidth]{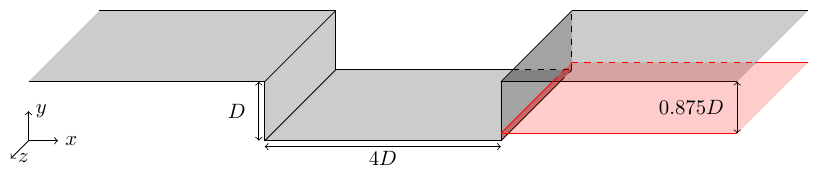}
\caption{The baseline, in black, and optimum, in red, designs of the open cavity.}
\label{fig_baseline_optimum_cavity_3d}
\end{figure}

\begin{figure}
\centering
\begin{subfigure}{0.7\textwidth}
\includegraphics[width=\textwidth]{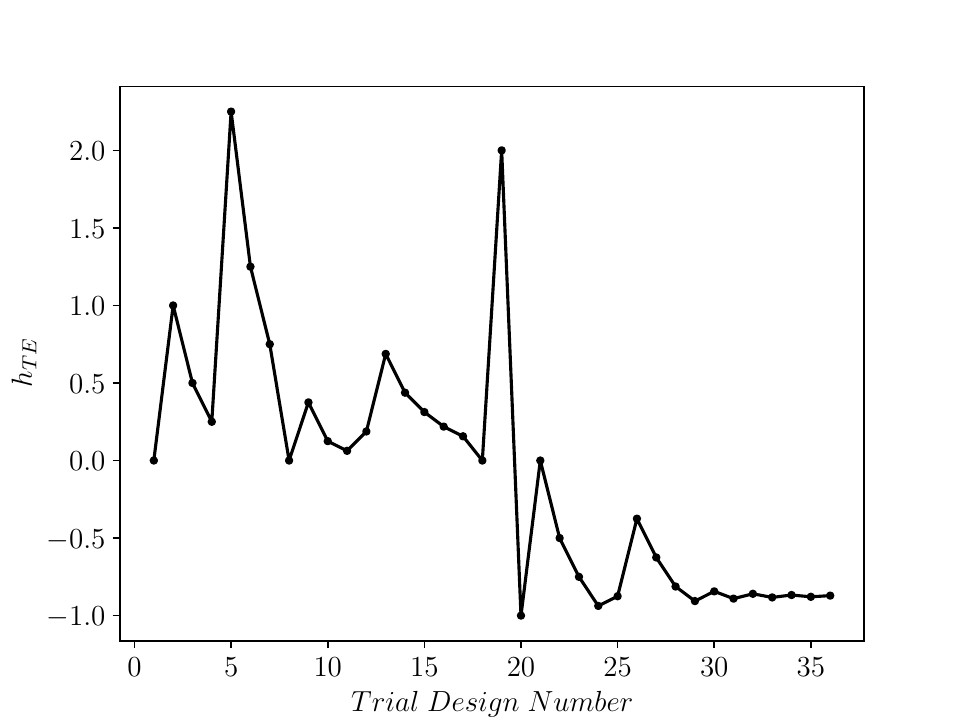}
\subcaption{The design space.}
\label{fig_cavity_designspace}
\end{subfigure}
\begin{subfigure}{0.7\textwidth}
\centering
\includegraphics[width=\textwidth]{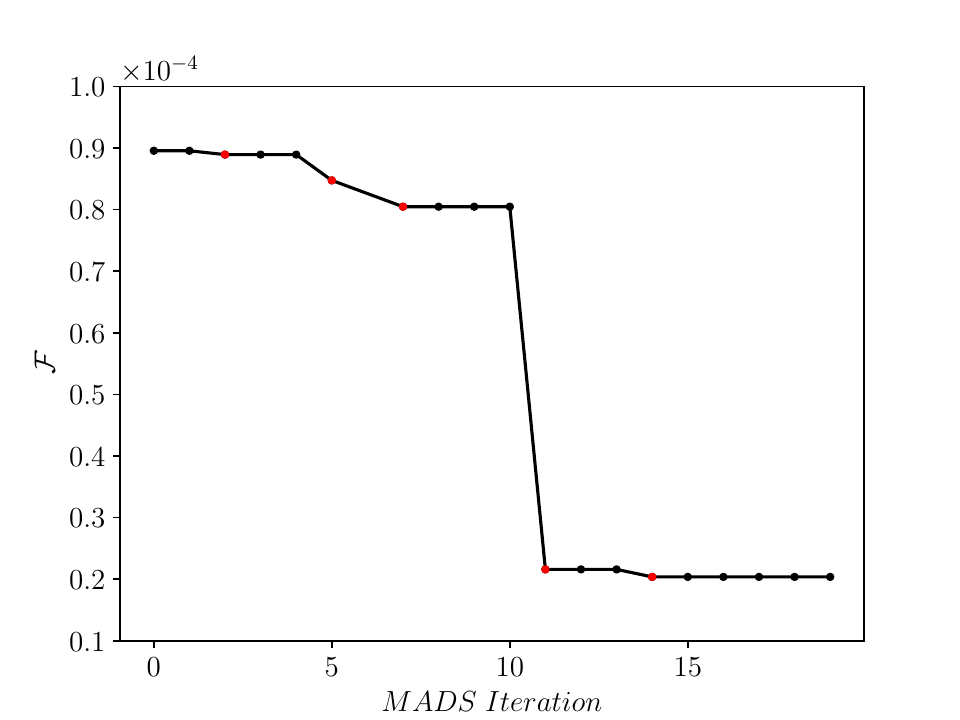}
\subcaption{The objective function convergence with the new incumbent designs
highlighted in red.}
\label{fig_cavity_objective}
\end{subfigure}
\caption{The design space and objective function convergence for the three-dimensional open deep cavity.}
\label{fig_design_space_and_obj_fun_cavity}
\end{figure}

The optimization procedure converged after $19$ MADS iterations with a total of $36$ objective function evaluations. The optimal design parameter is identified as $h_{TE}=-0.875$, resulting in an OASPL of $100.3 ~dB$, signifying a $13.0 ~dB$ reduction in noise. The baseline and optimum designs, depicted in Figure \ref{fig_baseline_optimum_cavity_3d}, illustrate a notable reduction in emitted noise by lowering the trailing edge wall of the cavity. However, such a modification may pose feasibility challenges in engineering applications. Moreover, Figure \ref{fig_design_space_and_obj_fun_cavity} illustrates the explored design parameter space and the convergence of the objective function.

\begin{figure}
\centering
\begin{subfigure}{\textwidth}
\centering
\includegraphics[width=\textwidth]{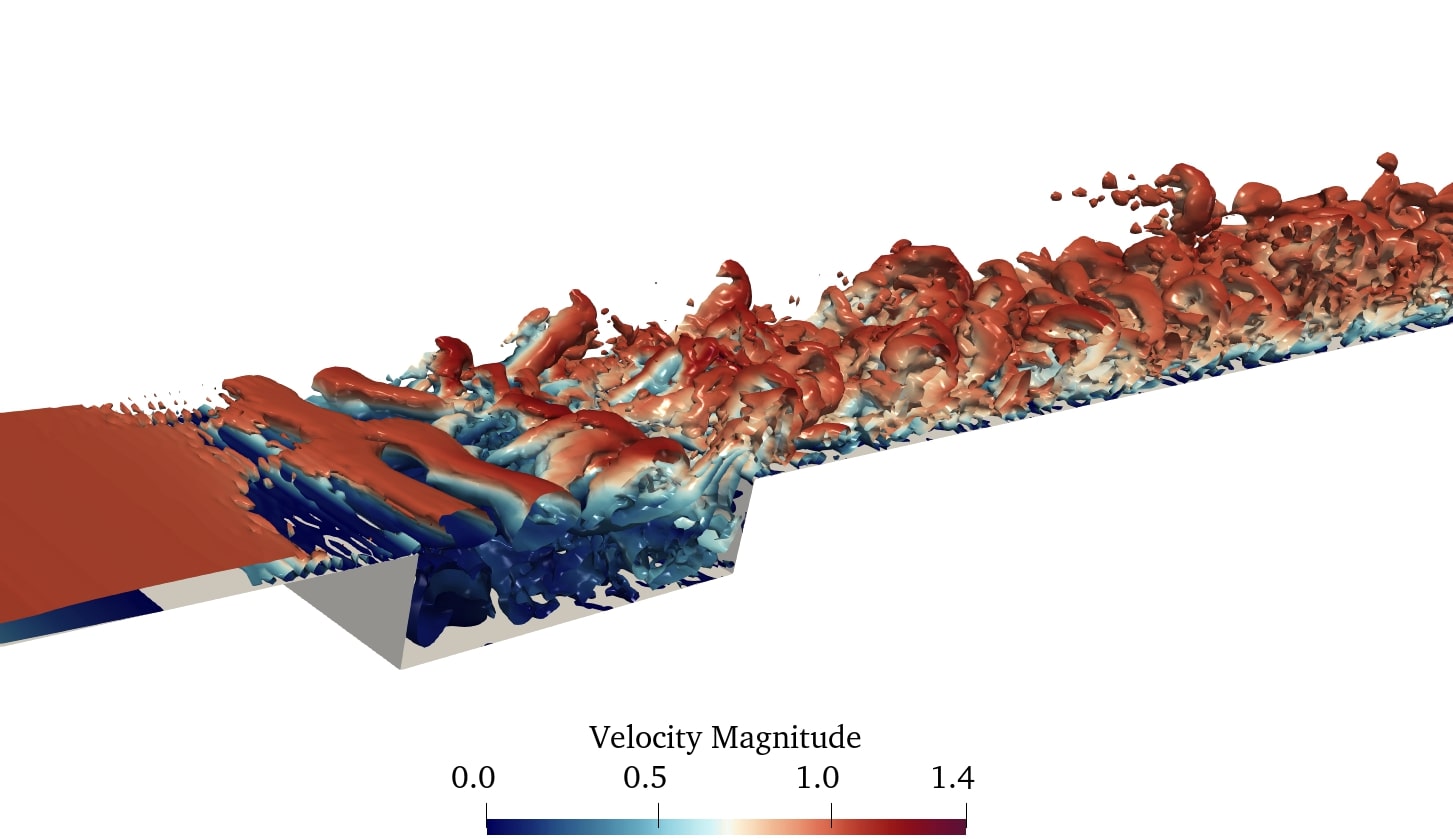}
\subcaption{Q-criterion contours coloured by velocity magnitude.}
\label{fig_cavity_3d_baseline_qcriterion}
\end{subfigure}
\begin{subfigure}{\textwidth}
\centering
\includegraphics[width=\textwidth]{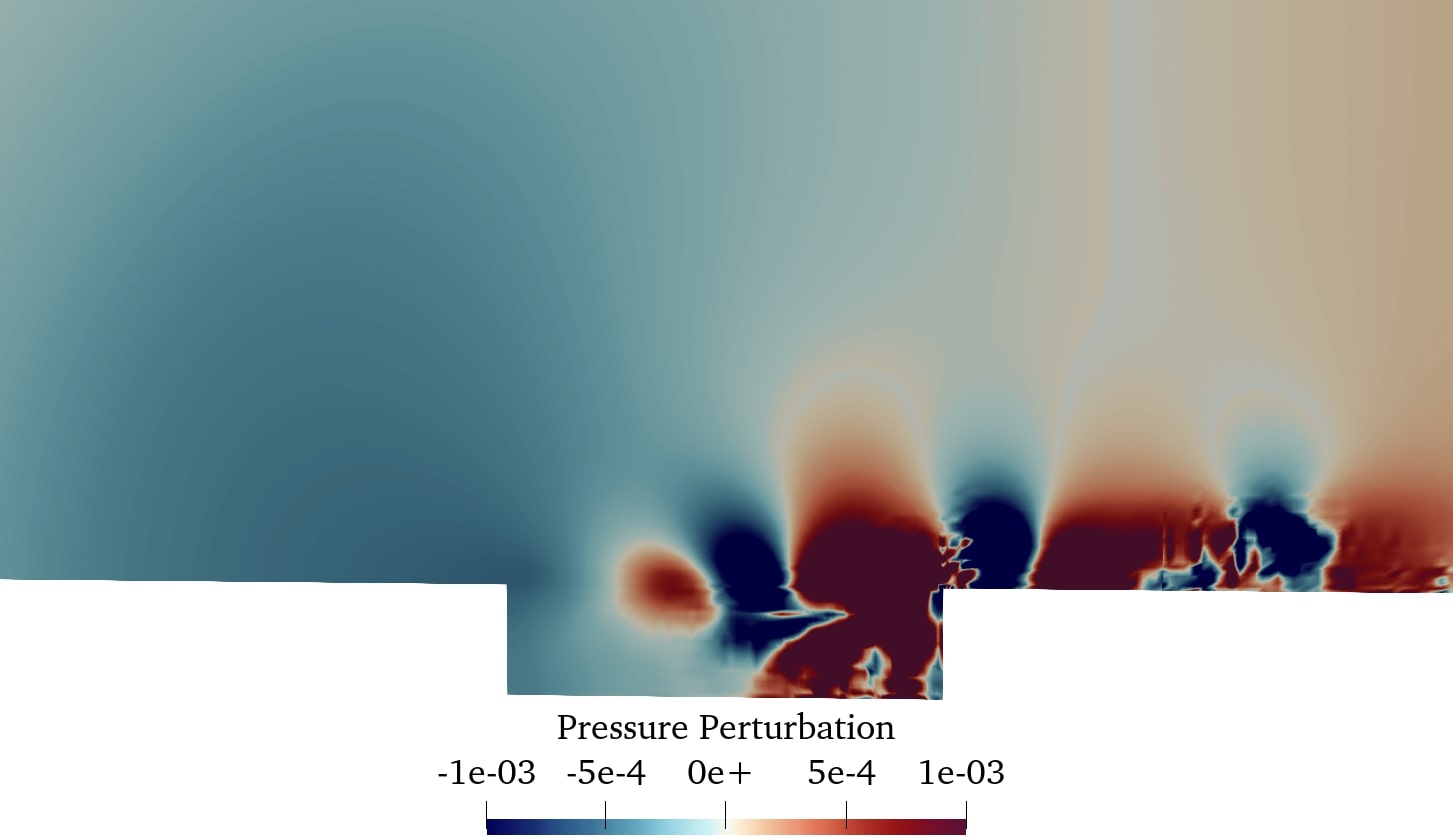}
\subcaption{Pressure perturbation.}
\label{fig_cavity_3d_baseline_acoustic}
\end{subfigure}
\caption{The Q-criterion contours and pressure perturbation for the baseline design of the open deep cavity.}
\label{fig_cavity_3d_baseline_qcriterion_acoustic}
\end{figure}

\begin{figure}
\centering
\begin{subfigure}{\textwidth}
\centering
\includegraphics[width=\textwidth]{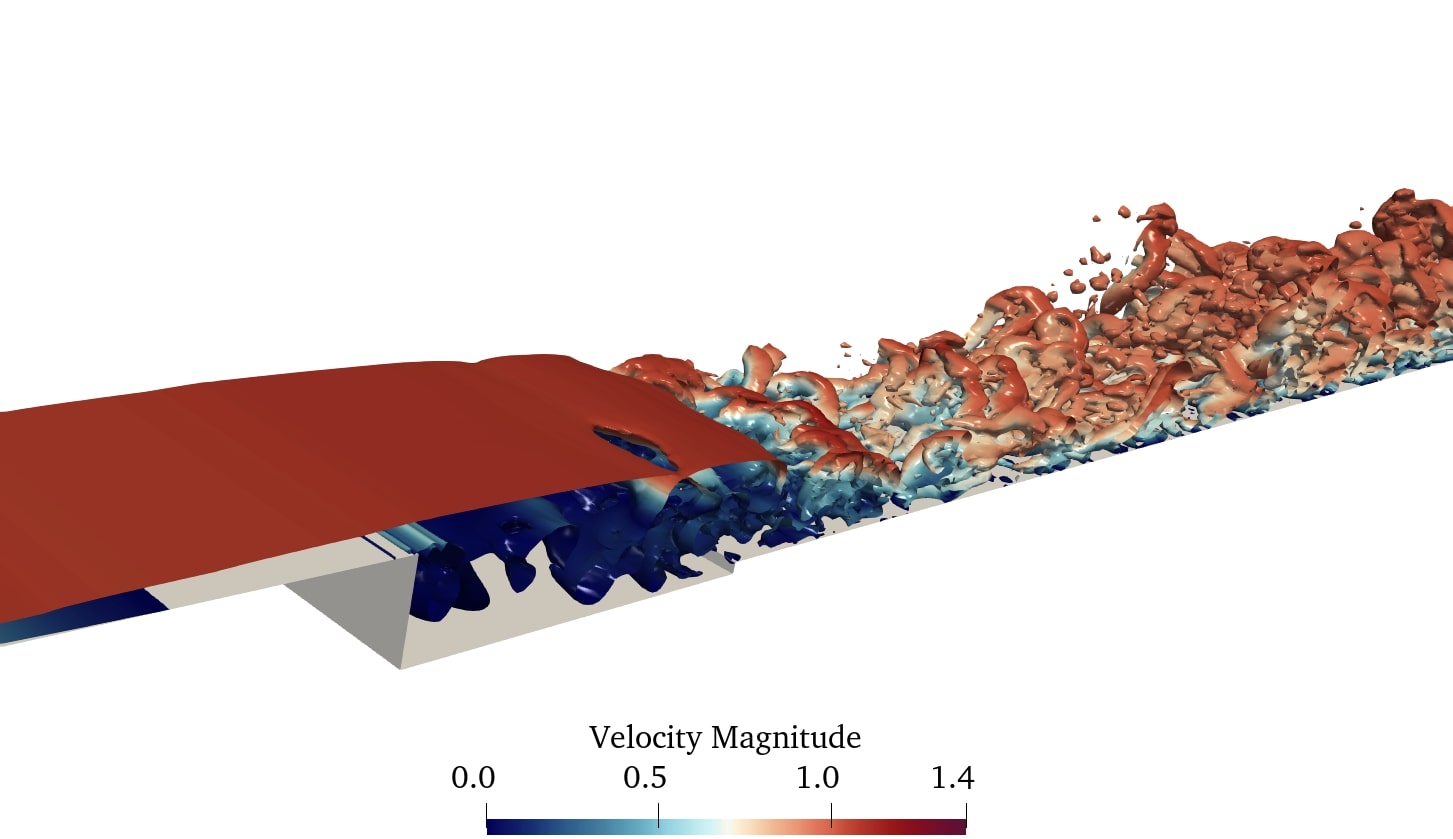}
\subcaption{Q-criterion contours coloured by velocity magnitude.}
\label{fig_cavity_3d_optimum_qcriterion}
\end{subfigure}
\begin{subfigure}{\textwidth}
\centering
\includegraphics[width=\textwidth]{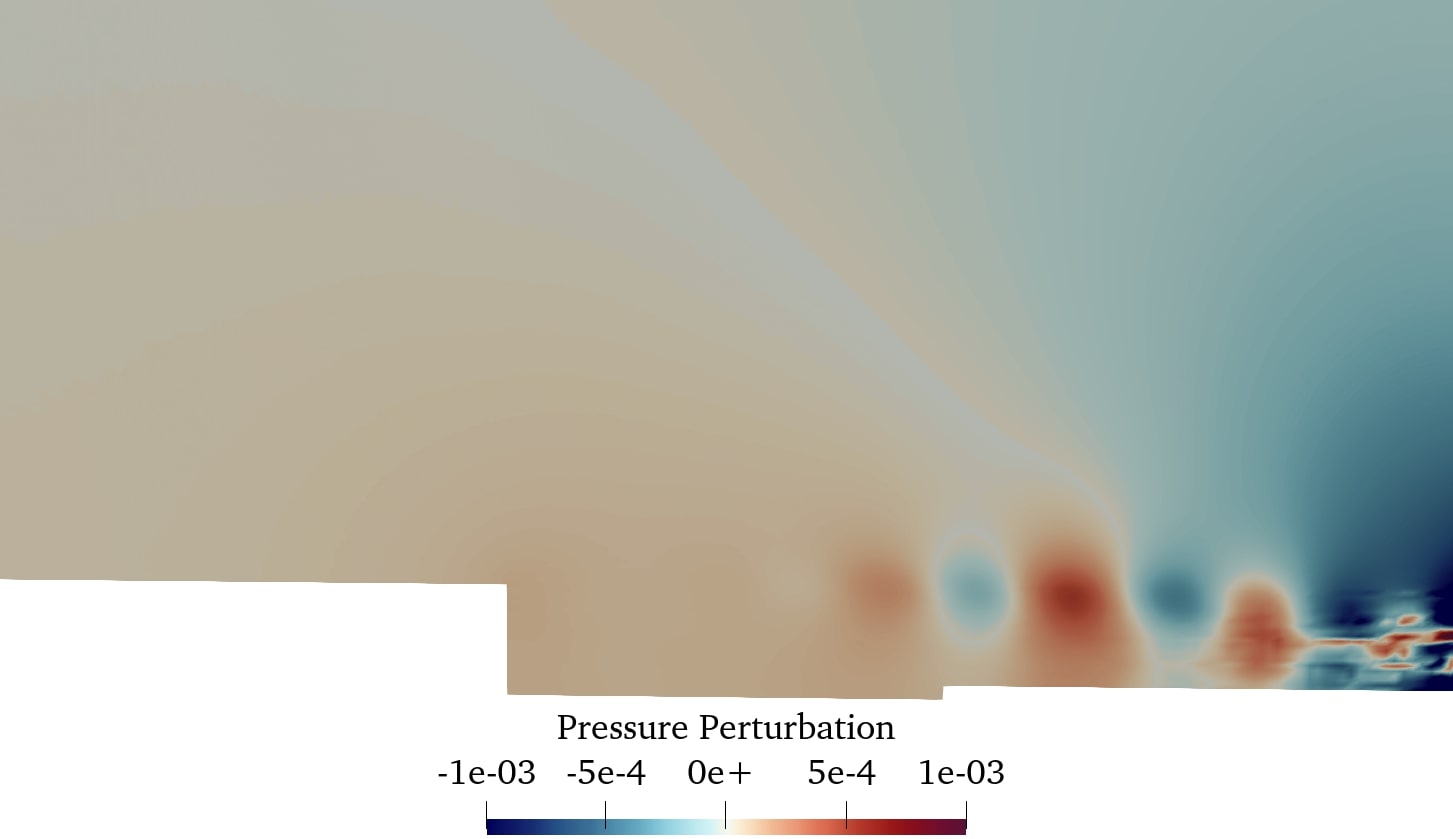}
\subcaption{Pressure perturbation.}
\label{fig_cavity_3d_optimum_acoustic}
\end{subfigure}
\caption{The Q-criterion contours and pressure perturbation for the optimum design of the open deep cavity.}
\label{fig_cavity_3d_optimum_qcriterion_acoustic}
\end{figure}

\begin{figure}
\centering
\includegraphics[width=0.7\textwidth]{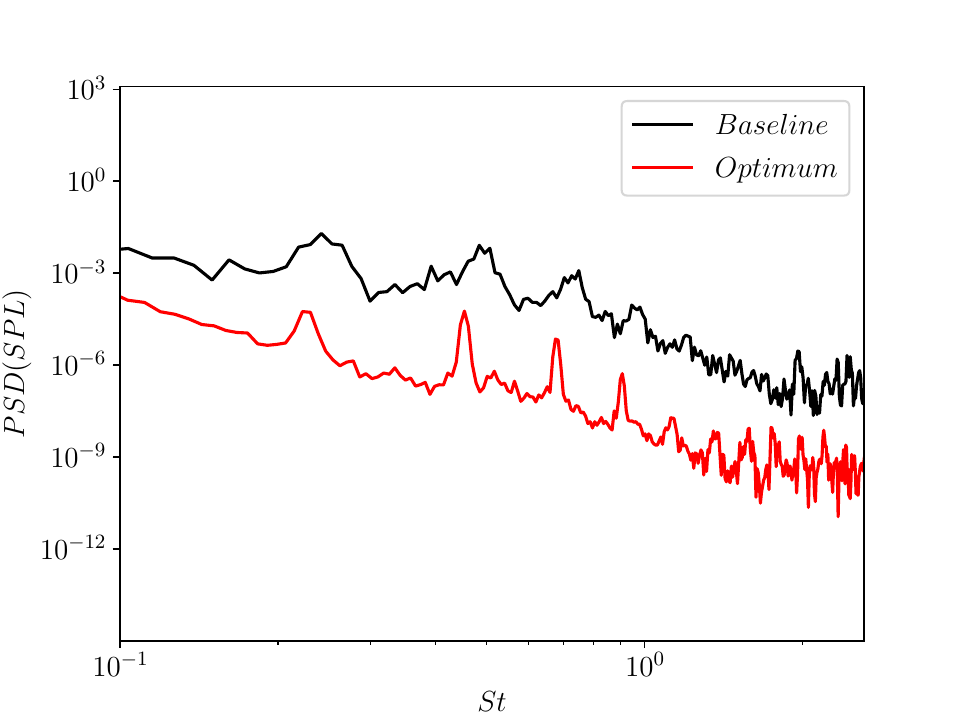}
\caption{The sound spectra for the open deep cavity.}
\label{fig_cavity_3d_sound_spectra}
\end{figure}

The Q-criterion contours coloured by velocity magnitude and the pressure perturbation of both the baseline and optimum designs are shown in Figures \ref{fig_cavity_3d_baseline_qcriterion_acoustic} and \ref{fig_cavity_3d_optimum_qcriterion_acoustic}, respectively. Comparing these figures, turbulent structures over the cavity are reduced significantly in the optimum design, and the shear layer expands over the cavity, resulting in much lower noise emission. Furthermore, the Power Spectral Density (PSD) of the OASPL is plotted against the Strouhal number for both the baseline and optimum designs in Figure \ref{fig_cavity_3d_sound_spectra}, which follows the Welch's method of periodiograms \cite{welch1967use} and involves dividing the time period into $6$ windows with a $50\%$ overlap. This figure illustrates the OASPL reduction across all frequency ranges.

\section{Tandem Cylinders}
\label{sec:TandemCylinders}

\begin{figure}
\centering
\includegraphics[width=0.7\textwidth]{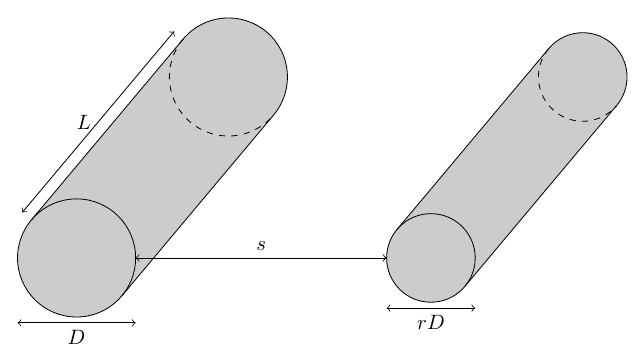}
\caption{The geometry of two cylinders in a tandem configuration.}
\label{fig_cylinder_geo}
\end{figure}

The flow around two tandem cylinders consists of multiple flow features including flow separation, reattachment, recirculation, and quasi-periodic vortex shedding, amongst others. The physics of such flows is highly dependent on the diameter ratio of the cylinders, the spacing between them, and the Reynolds number. The diameter ratio of the cylinders is defined as $r=D_{d}/D_{u}$, where $D_d$ and $D_u$ are the downstream and upstream diameter of the cylinders, respectively. The spacing of the cylinders, $s$, is defined as the distance between the rear of the upstream cylinder to the front of the downstream cylinder. These definitions are depicted in Figure \ref{fig_cylinder_geo}.

The three-dimensional wake development of a single cylinder was studied by Williamson \cite{williamson1996vortex}. Additionally, Papaioannou et al. \cite{papaioannou2006three} investigated the three-dimensionality effects of flow over two tandem cylinders, varying Reynolds number and the spacing distance between the cylinders. They found that as Reynolds number increased, two-dimensional results diverged from three-dimensional ones, especially beyond a critical Reynolds number where wake three-dimensionality initiated. The Reynolds number of our study, based on the upstream cylinder's diameter, is $Re_D=1000$ since the wake will develop considerable three-dimensionality and this Reynolds number is associated with the early turbulent regime \cite{papaioannou2006three}.

\subsection{Validation}

In this section, the simulation of flow over two tandem cylinders is validated using reference DNS data \cite{papaioannou2006three}, along with grid independence study of the time-averaged lift and drag coefficients and OASPL at a near-field observer located $2D$ above the upstream cylinder. Then, the optimization is performed similar to our previous work \cite{hamedi2024near}, where sound at the near-field observer is minimized. The design variables are the ratio of the cylinders' diameters, $r$, and the distance between the two, $s$.

\subsubsection{Computational Details}

\begin{figure}
\centering
\includegraphics[width=0.7\textwidth]{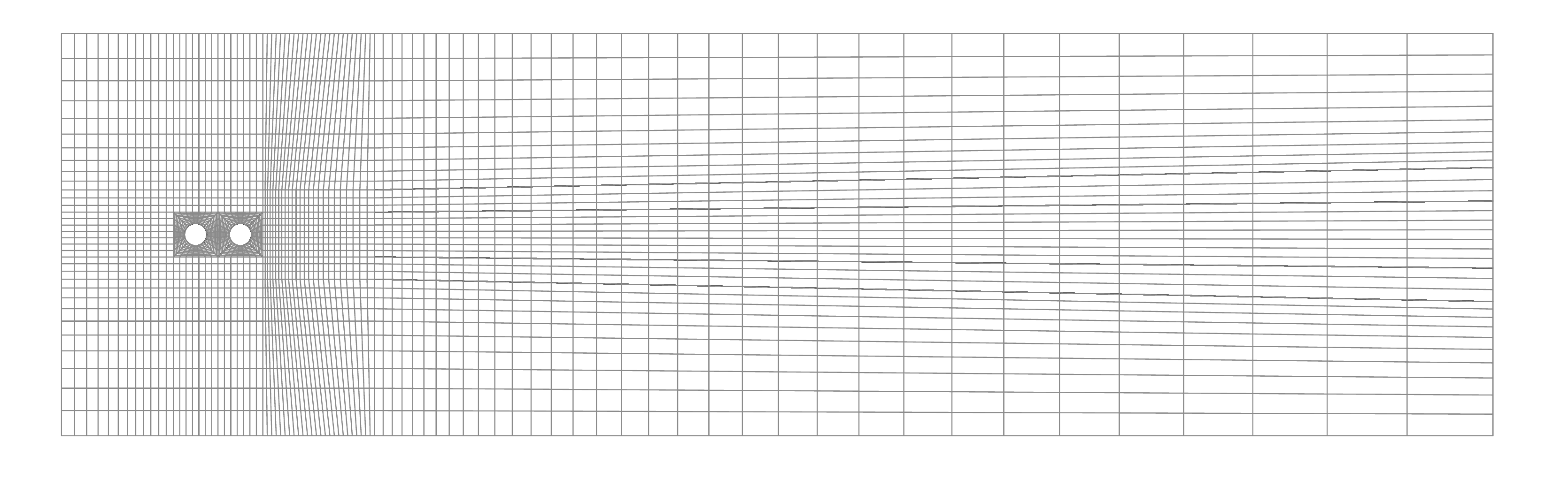}
\caption{The mesh of the two cylinders in a tandem configuration.}
\label{fig_cylinders_mesh_3d}
\end{figure}

The cylinders are located at a distance of $s/D=1$ with a ratio of $r=1$ and have a spanwise length of $L/D=10$, following previous studies \cite{papaioannou2006three}. The Reynolds number, based on the upstream cylinder's diameter, is $Re_D=1000$, corresponding to the early turbulent regimes \cite{papaioannou2006three}, and the Mach number is $0.2$. The boundary layer region extends to $0.5D$ around the cylinders, with the inlet boundary placed $5D$ away from the upstream cylinder and the outlet boundary $55D$ away from the downstream cylinder. The computational domain is extended to $10D$ in the $y$-direction. The stretching ratio for the first $5D$ and $1D$ elements in the $x$ and $y$-directions, respectively, is $1.05$, and that of the remaining elements is $1.075$. The smallest element size in the domain is $0.1D$, which is in the boundary layer region. A total number of $31,780$ hexahedral elements are used. The mesh of the tandem cylinders is shown in Figure \ref{fig_cylinders_mesh_3d}. Periodic boundary conditions are applied in the spanwise direction, while a no-slip boundary condition is imposed on the surface of the cylinders, along with Riemann invariant boundary conditions at the inlet and outlet of the computational domain. The simulation is run for $100t_c$, where $t_c=D/U_\infty$, to allow initial transients to disappear, followed by a subsequent period of $500t_c$ to obtain an average of the statistical quantities.

\subsubsection{Results and Discussion}

\begin{figure}
\centering
\includegraphics[width=0.7\textwidth]{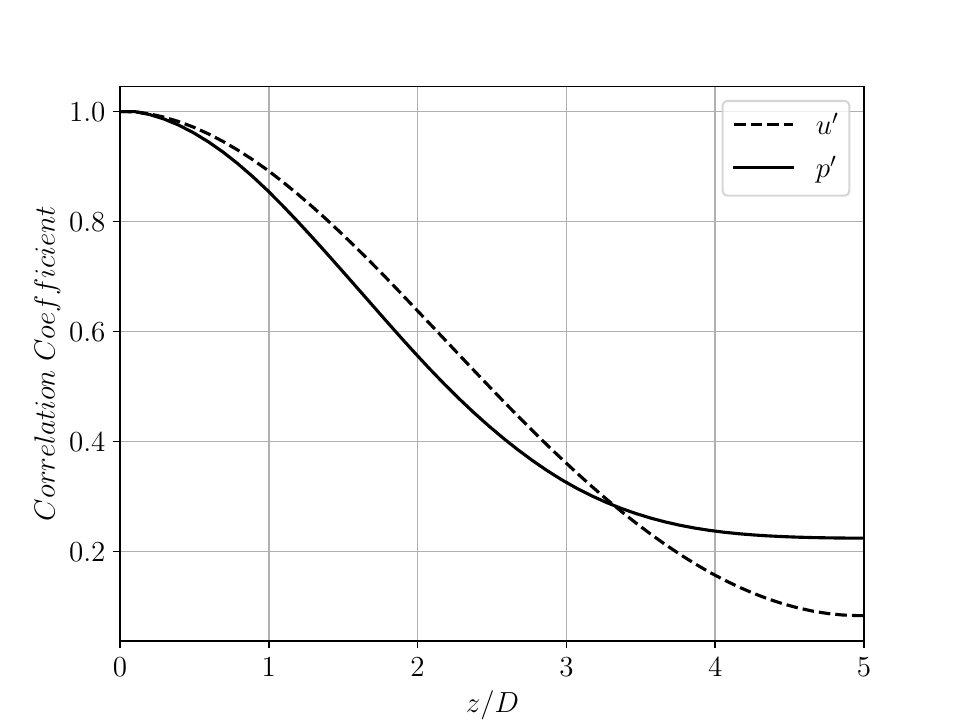}
\caption{The correlation coefficient in the spanwise direction for the tandem cylinders.}
\label{fig_cylinder_correlation}
\end{figure}

The sufficiency of the spanwise length is investigated by computing the correlation coefficient of the velocity fluctuation and the pressure perturbation along the $z$-direction. The correlation plot, demonstrated in Figure \ref{fig_cylinder_correlation}, ensures the uncorrelated fluctuations in the $z$-direction at a separation of half of the domain size.
Furthermore, the time-averaged drag coefficient and the OASPL at the observer are computed using different averaging window lengths, summarized in Table \ref{table_cylinders_averaging_window}. The time-averaged drag coefficient of the upstream cylinder is computed using $\mathcal{P}2$ and $\mathcal{P}3$ simulations. The $\overline{C_D}_1$ obtained using the $\mathcal{P}3$ simulation is $0.997$, which is in good agreement with the reference value of $0.988$ \cite{papaioannou2006three}. Table \ref{table_cylinders_averaging_window} shows that the difference in the statistical time-averaged quantities is negligible beyond $500 t_c$. Thus, in this study, the statistical quantities are averaged for $500t_c$.
\begin{table}
\centering
\caption{The $\overline{C_D}_1$ and OASPL at the observer, for the tandem cylinders configuration using different lengths of the averaging window.}
\begin{tabular}{ccccc}
\hline
Averaging Window Size & \multicolumn{2}{c}{$\overline{C_D}_1$} & \multicolumn{2}{c}{OASPL in $dB$} \\
\hline
 & $\mathcal{P}2$ & $\mathcal{P}3$ & $\mathcal{P}2$ & $\mathcal{P}3$ \\
\hline
$200t_c$ & $0.962374$ & $0.994465$ & $126.5$ & $125.1$ \\
$300t_c$ & $0.963871$ & $0.994915$ & $126.9$ & $125.2$ \\
$400t_c$ & $0.965569$ & $0.996092$ & $127.3$ & $125.2$ \\
$500t_c$ & $0.966651$ & $0.996752$ & $127.6$ & $125.2$ \\
$600t_c$ & $0.967519$ & $0.997042$ & $127.7$ & $125.3$ \\
$700t_c$ & $0.968142$ & $0.996965$ & $127.8$ & $125.3$ \\
%$800t_c$ & $0.968674$ & $0.996713$ & $127.918$ & $125.299$ \\
%$900t_c$ & $0.968925$ & $0.997093$ & $127.973$ & $125.283$ \\
\hline
\end{tabular}
\label{table_cylinders_averaging_window}
\end{table}

\subsection{Optimization}

The distance between the two cylinders, $s$, and the ratio between the diameters of the cylinders, $r$, are the design variables, $\pmb{\mathcal{X}} = [s, r]$. The objective function is $\mathcal{F} = p^\prime_{rms}$ at $2D$ above the upstream cylinder.

\subsubsection{Results and Discussion}

\begin{figure}
\centering
\includegraphics[width=0.7\textwidth]{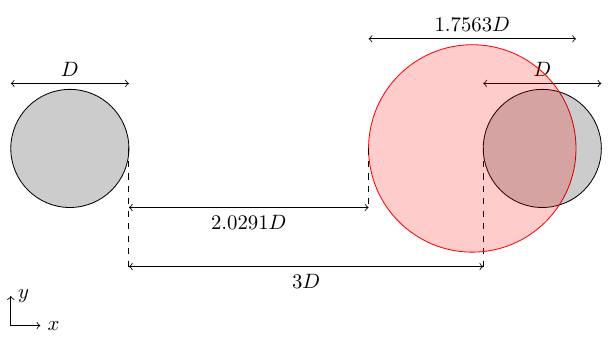}
\caption{The baseline, in black, and optimum, in red, designs of the tandem cylinders.}
\label{fig_baseline_optimum_cylinders_3d}
\end{figure}

\begin{figure}
\centering
\begin{subfigure}{0.7\textwidth}
\includegraphics[width=\textwidth]{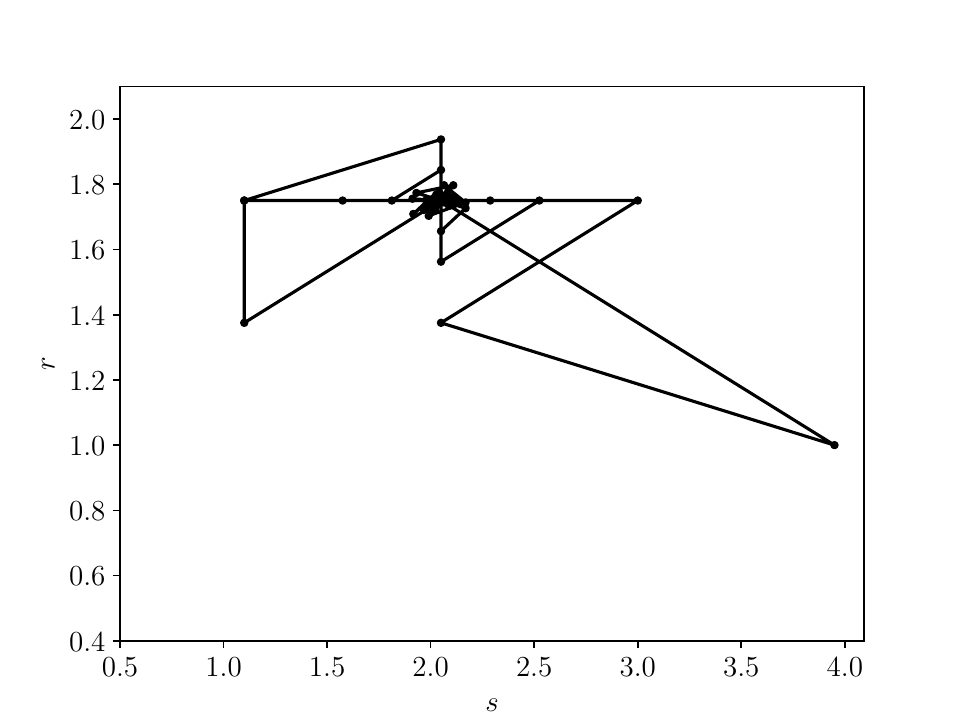}
\subcaption{The design space.}
\label{figure_cylinder_optimization_3d_design_space}
\end{subfigure}
\begin{subfigure}{0.7\textwidth}
\includegraphics[width=\textwidth]{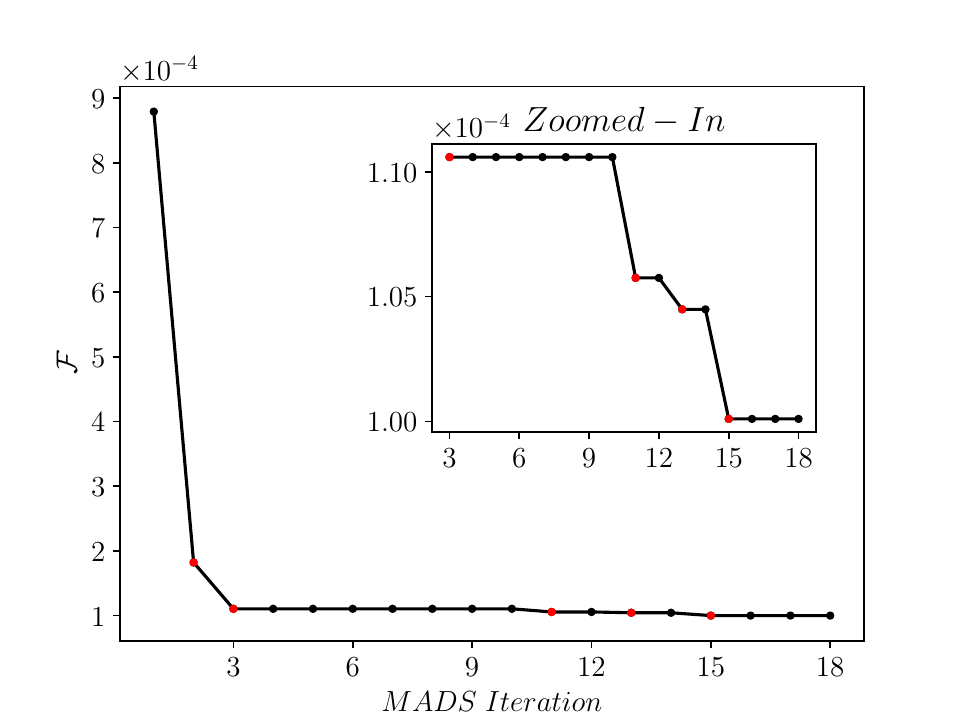}
\subcaption{The objective function convergence with the new incumbent designs highlighted in red.}
\label{figure_cylinder_optimization_3d_obj}
\end{subfigure}
\caption{The design space and objective function convergence for the tandem cylinders.}
\label{figure_cylinder_optimization_3d}
\end{figure}

\begin{figure}
\centering
\begin{subfigure}{\textwidth}
\centering
\includegraphics[width=\textwidth]{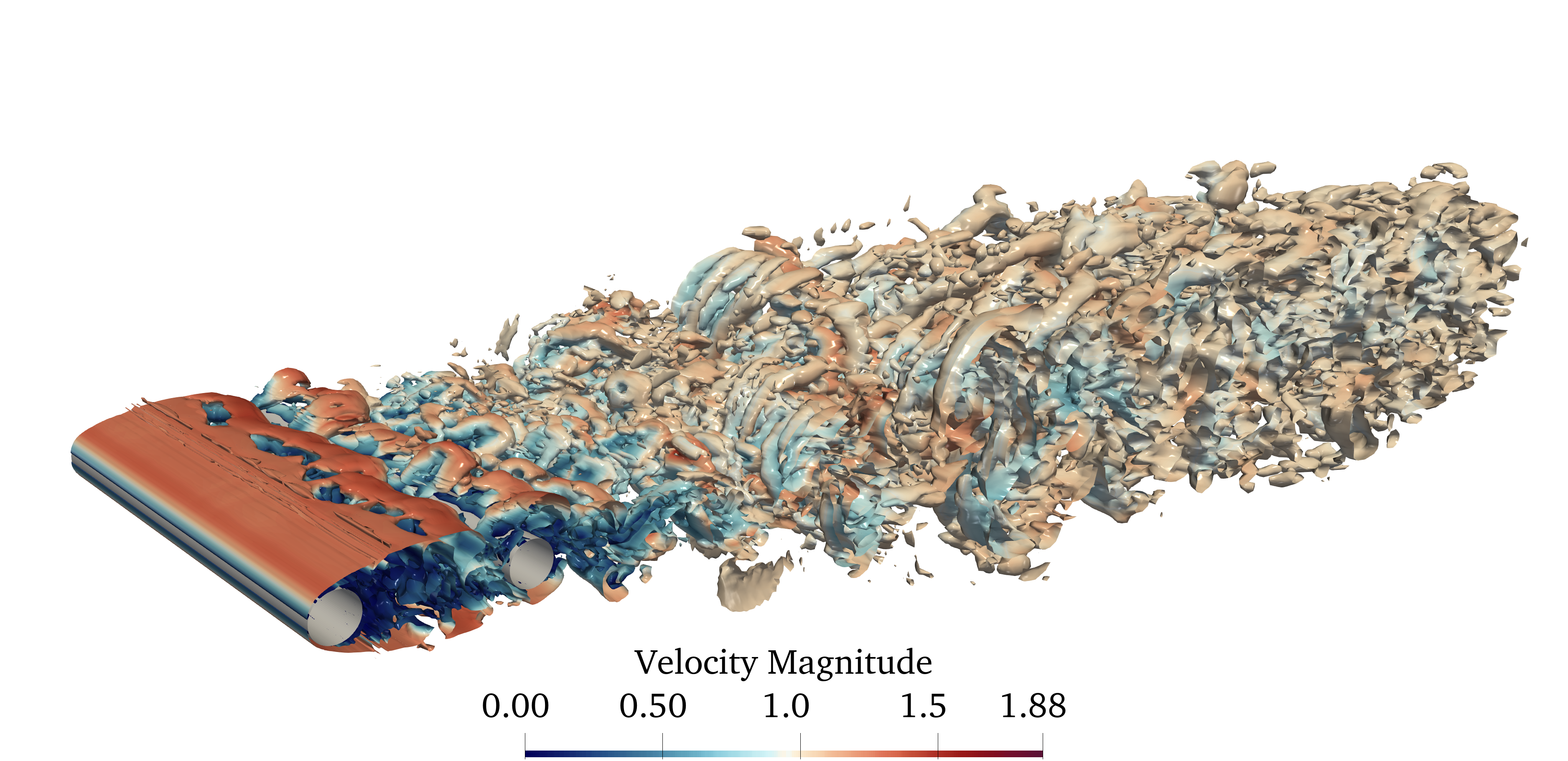}
\subcaption{Q-criterion coloured by velocity magnitude.}
\end{subfigure}
\begin{subfigure}{\textwidth}
\centering
\includegraphics[width=0.7\textwidth]{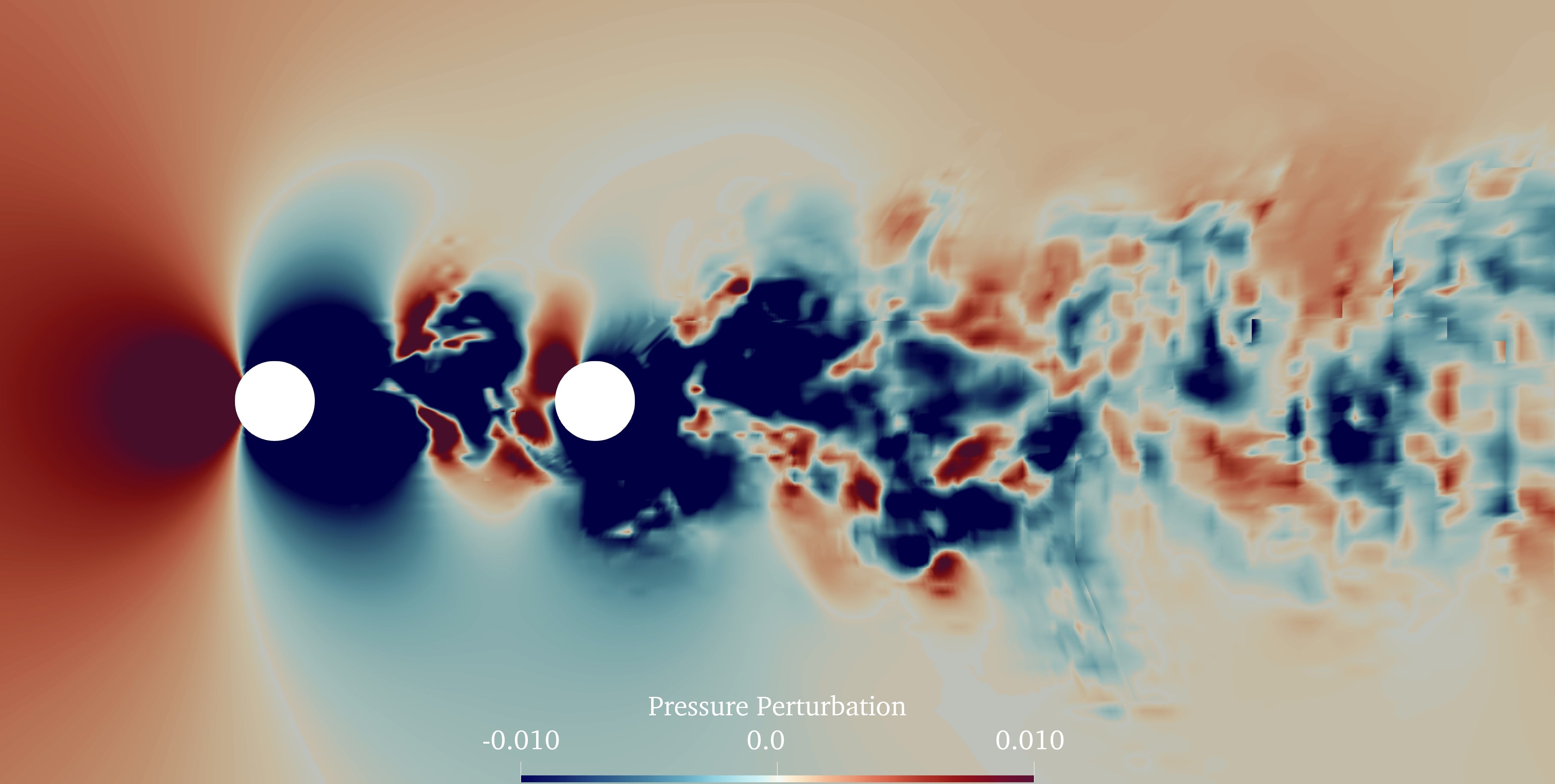}
\subcaption{Acoustic pressure field at mid-plane.}
\end{subfigure}
\caption{The baseline tandem cylinder design at $t_c=600$.}
\label{fig:Cylinders3DinitialDesign}
\end{figure}

\begin{figure}
\centering
\begin{subfigure}{\textwidth}
\centering
\includegraphics[width=\textwidth]{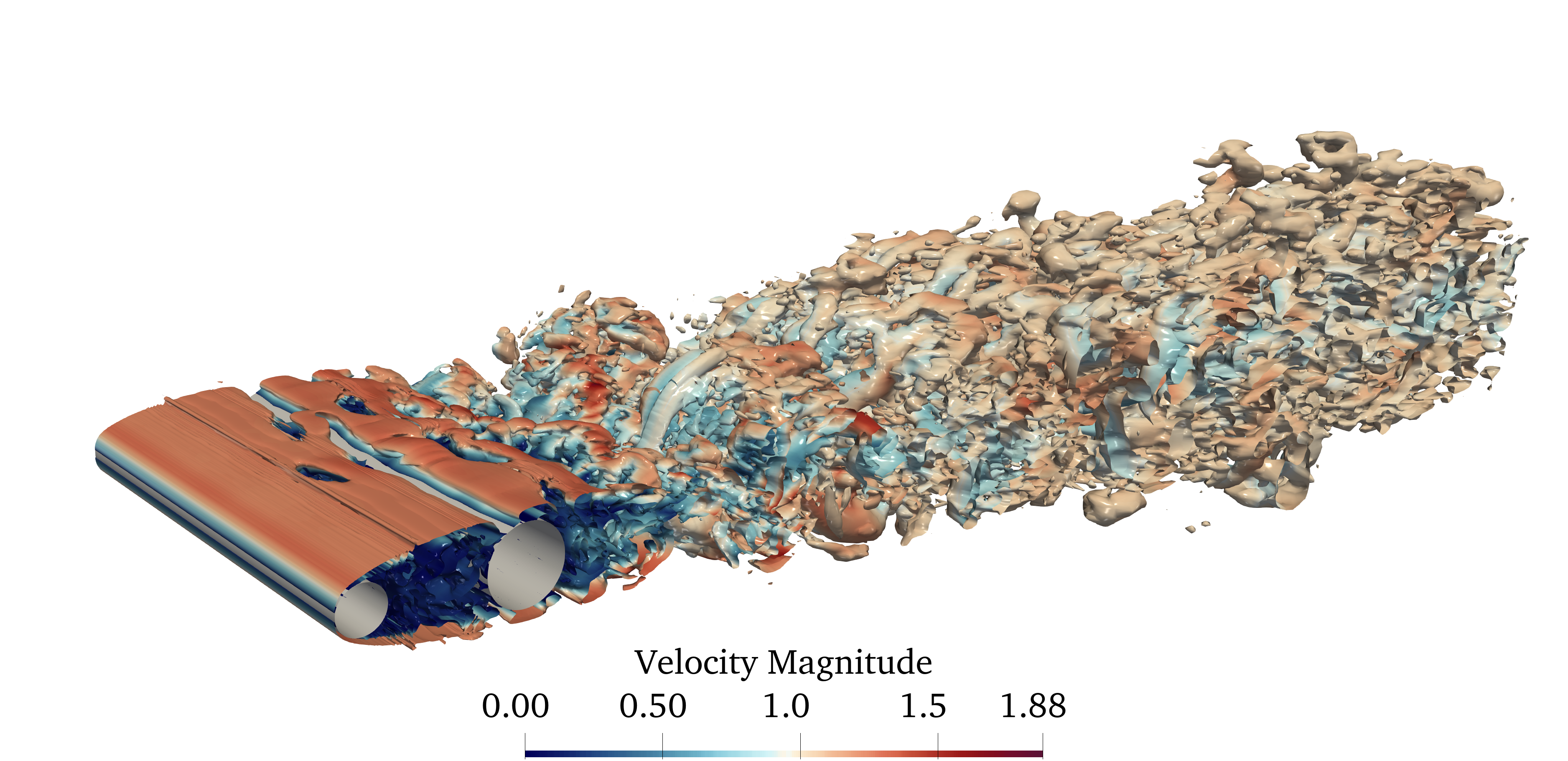}
\subcaption{Q-criterion coloured by velocity magnitude.}
\end{subfigure}
\begin{subfigure}{\textwidth}
\centering
\includegraphics[width=0.7\textwidth]{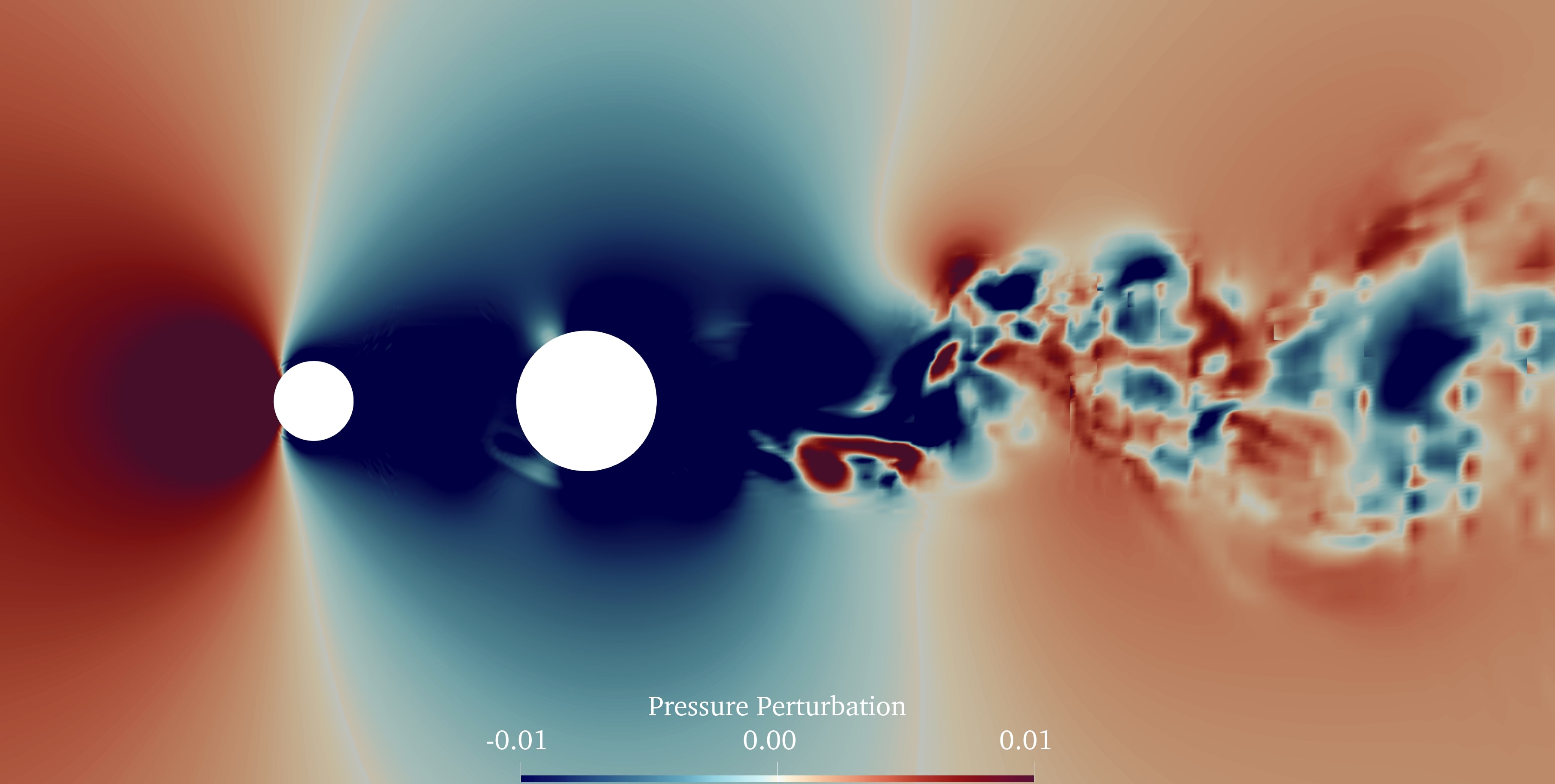}
\subcaption{Acoustic pressure field at mid-plane.}
\end{subfigure}
\caption{The optimum tandem cylinder design at $t_c=600$.}
\label{fig:Cylinders3DoptimumDesign}
\end{figure}

\begin{figure}
\centering
\includegraphics[width=0.7\textwidth]{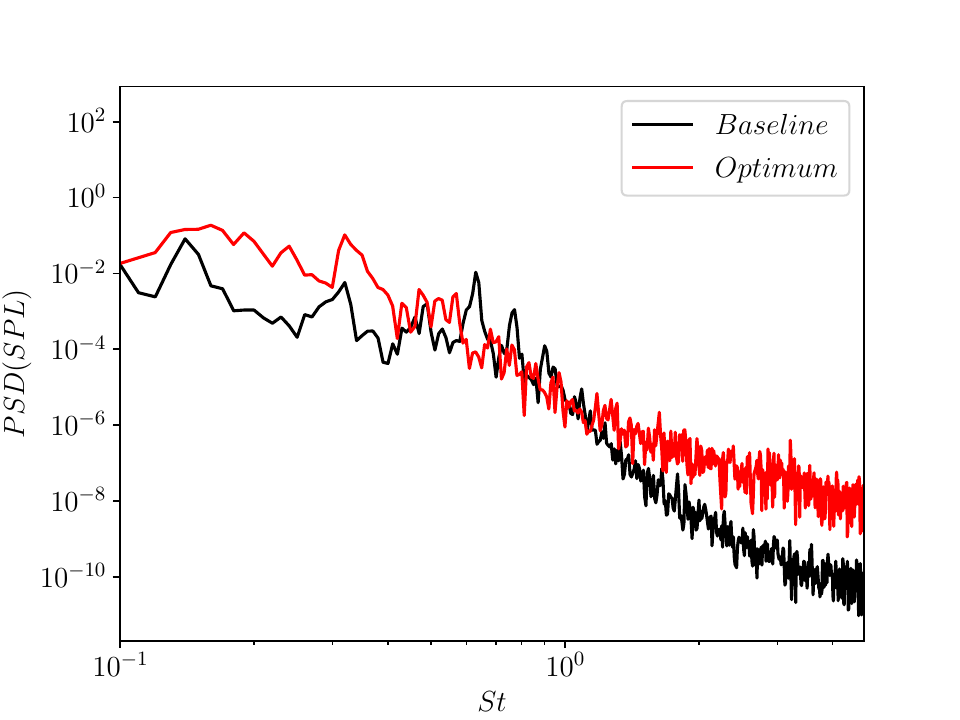}
\caption{The sound spectra for the tandem cylinders.}
\label{fig_cylinders_3d_sound_spectra}
\end{figure}

The optimization problem converges in $18$ MADS iterations with a total of $48$ objective function evaluations. The baseline and optimum designs are shown in Figure \ref{fig_baseline_optimum_cylinders_3d}. The design space and objective function convergence are shown in Figure \ref{figure_cylinder_optimization_3d}, where the optimum design is found as $(s,r)=(2.0291D,1.7563D)$. The optimization process explores a wide range of design variables, as illustrated in Figure \ref{figure_cylinder_optimization_3d_design_space}. Q-criterion contours, coloured by velocity magnitude, and acoustic field at the mid-plane are shown for both the baseline and optimum designs in Figures \ref{fig:Cylinders3DinitialDesign} and \ref{fig:Cylinders3DoptimumDesign}, respectively. The optimized design exhibits a smoother flow field, resulting in reduced noise emissions. The OASPL of the initial design at the observer, $2D$ above the upstream cylinder, is $125.3~dB$, which decreases to $114.1~dB$ for the optimized configuration. Lastly, Figure \ref{fig_cylinders_3d_sound_spectra} presents the PSD of OASPL versus Strouhal number, computed using Welch's method of periodograms \cite{welch1967use} with $3$ windows and a $50\%$ overlap. It is evident that the optimum design displays higher intensity PSD of OASPL over a broad frequency range, while achieving a lower OASPL value, primarily due to a decrease in the largest magnitude modes. Furthermore, this behavior can be attributed to the baseline design producing high-intensity sound at specific frequencies ($St=0.63, 0.77,$ and $0.90$), contributing to its elevated peak OASPL, whereas the optimum design distributes its energy across a wider frequency spectrum.

\section{NACA 4-digit Airfoil}
\label{sec:NACA}

The flow over NACA 4-digit airfoils is investigated in this section. The computational domain, previously used by the authors \cite{hamedi2024near}, is extruded in the $z$-direction. The validation of the flow simulation is conducted using an ILES reference \cite{kojima2013large} and a grid independence study for a NACA0012 airfoil. Subsequently, four design parameters, akin to those in \cite{hamedi2024near}, are selected, and the gradient-free MADS optimization technique is employed. The maximal positive basis construction is employed for the optimization algorithm.

\subsection{Validation} 

Validation for flow over a NACA0012 airfoil at an angle of attack of $6^\circ$ is conducted. The validation process involves comparing the time-averaged lift and drag coefficients obtained from two distinct grid resolutions with those from an ILES reference \cite{kojima2013large}. Moreover, the time-averaged pressure coefficient, the skin friction coefficient, and the OASPL at a near-field observer is computed using both grid resolutions and various time averaging window lengths. This analysis ensures the independence of the results to both grid resolution and time averaging window lengths. Detailed computational procedures and validation results are presented in the subsequent sections.

\subsubsection{Computational Details}

\begin{figure}
\centering
\begin{subfigure}{0.45\textwidth}
\centering
\includegraphics[width=\textwidth]{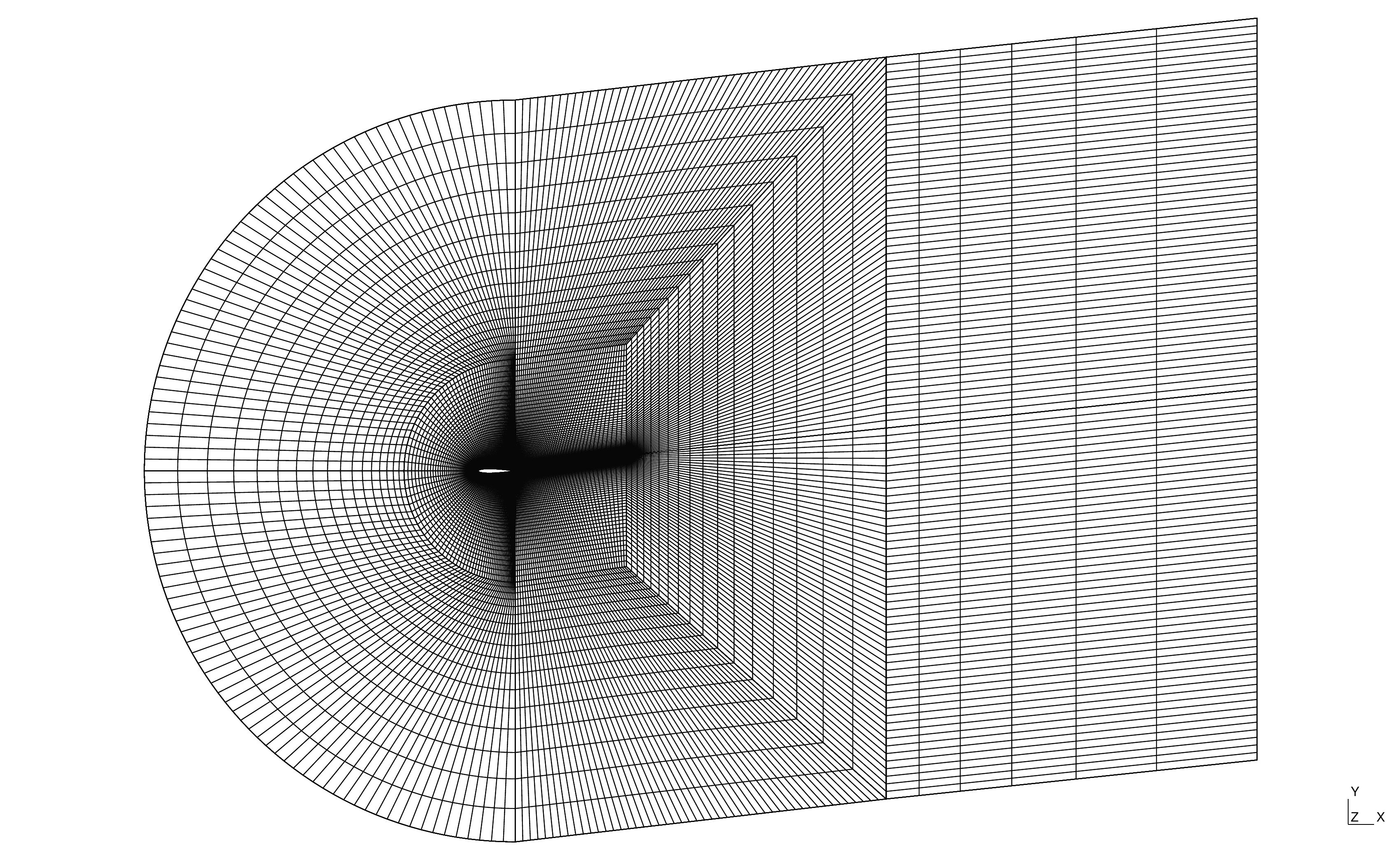}
\subcaption{The computational domain.}
\end{subfigure}
\begin{subfigure}{0.45\textwidth}
\centering
\includegraphics[width=\textwidth]{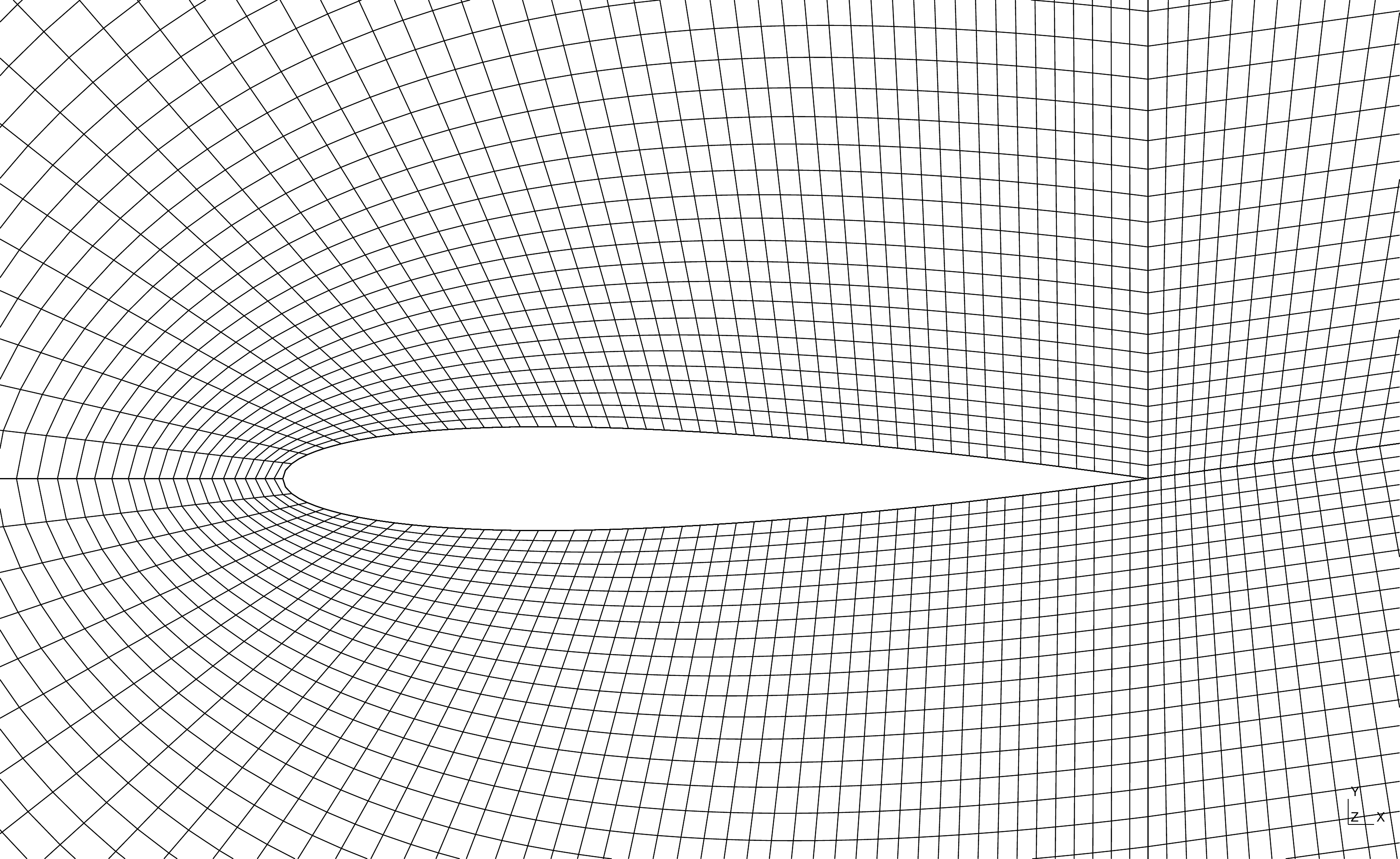}
\subcaption{The vicinity of the airfoil.}
\end{subfigure}
\caption{The computational grid for NACA0012 airfoil at $\alpha = 6^\circ$.}
\label{fig_naca_3d_domain}
\end{figure}

\begin{figure}
\centering
\begin{subfigure}{0.45\textwidth}
\centering
\includegraphics[width=\textwidth]{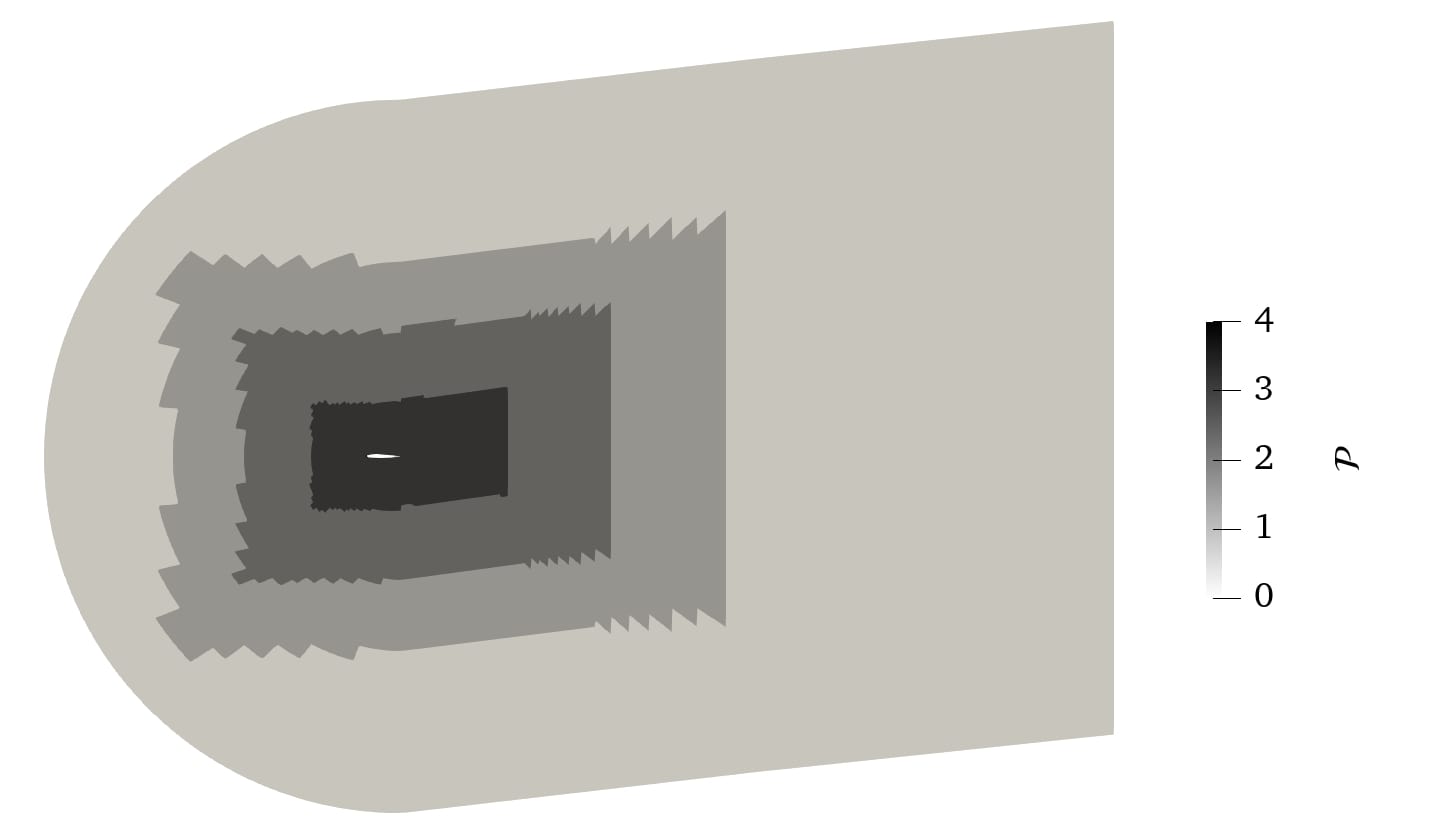}
\subcaption{Low resolution, $\mathcal{P}0 - \mathcal{P}3$.}
\label{fig_naca_3d_p_distribution_p3}
\end{subfigure}
\begin{subfigure}{0.45\textwidth}
\centering
\includegraphics[width=\textwidth]{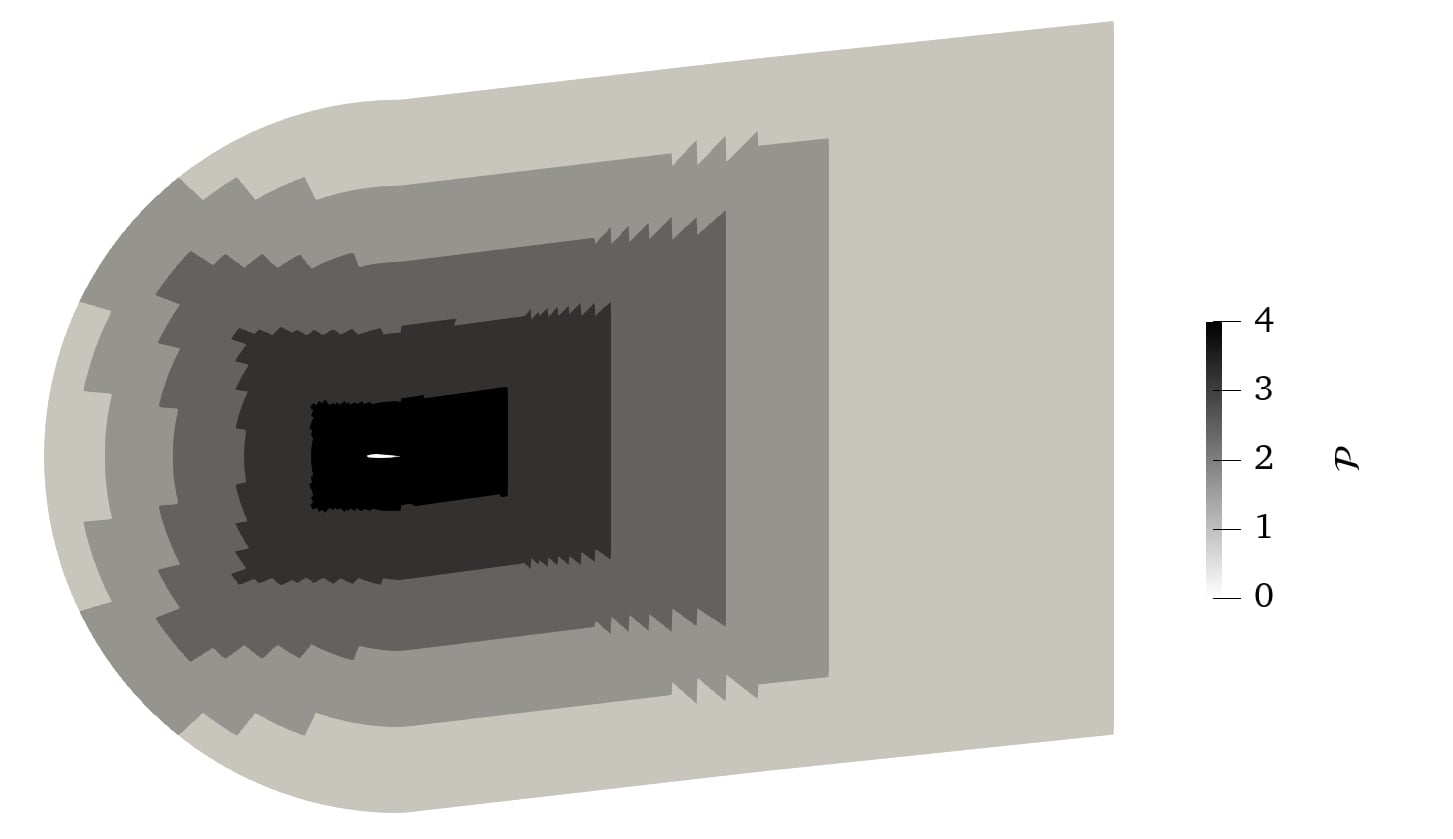}
\subcaption{High resolution, $\mathcal{P}0 - \mathcal{P}4$.}
\end{subfigure}
\caption{Different solution polynomial distributions for grid independence study of NACA0012 airfoil at $\alpha = 6^\circ$.}
\label{fig_naca_3d_p_distribution}
\end{figure}

The computational grid consists of $121,520$ hexahedral elements, illustrated in Figure \ref{fig_naca_3d_domain}. The domain extends to $20c$ in the $x$-direction, $10c$ in the $y$-direction, and $0.2c$ in the $z$-direction, with $c=1$ representing the airfoil chord. Notably, elements in the wake region are inclined at the angle of attack to accurately capture trailing-edge vortices. The flow conditions are characterized by a Reynolds number of $23,000$, a free-stream Mach number of $M=0.2$, and Prandtl number is $Pr=0.71$. The simulation is run for $10$ convective times to allow the initial transition disappears and then run for another $70$ convective times for flow statistics averaging. Additionally, a variable solution polynomial degree is implemented to eliminate acoustic wave reflections from boundaries, as demonstrated in Figure \ref{fig_naca_3d_p_distribution}.

\subsubsection{Results and Discussion}

\begin{figure}
\centering
\includegraphics[width=0.7\textwidth]{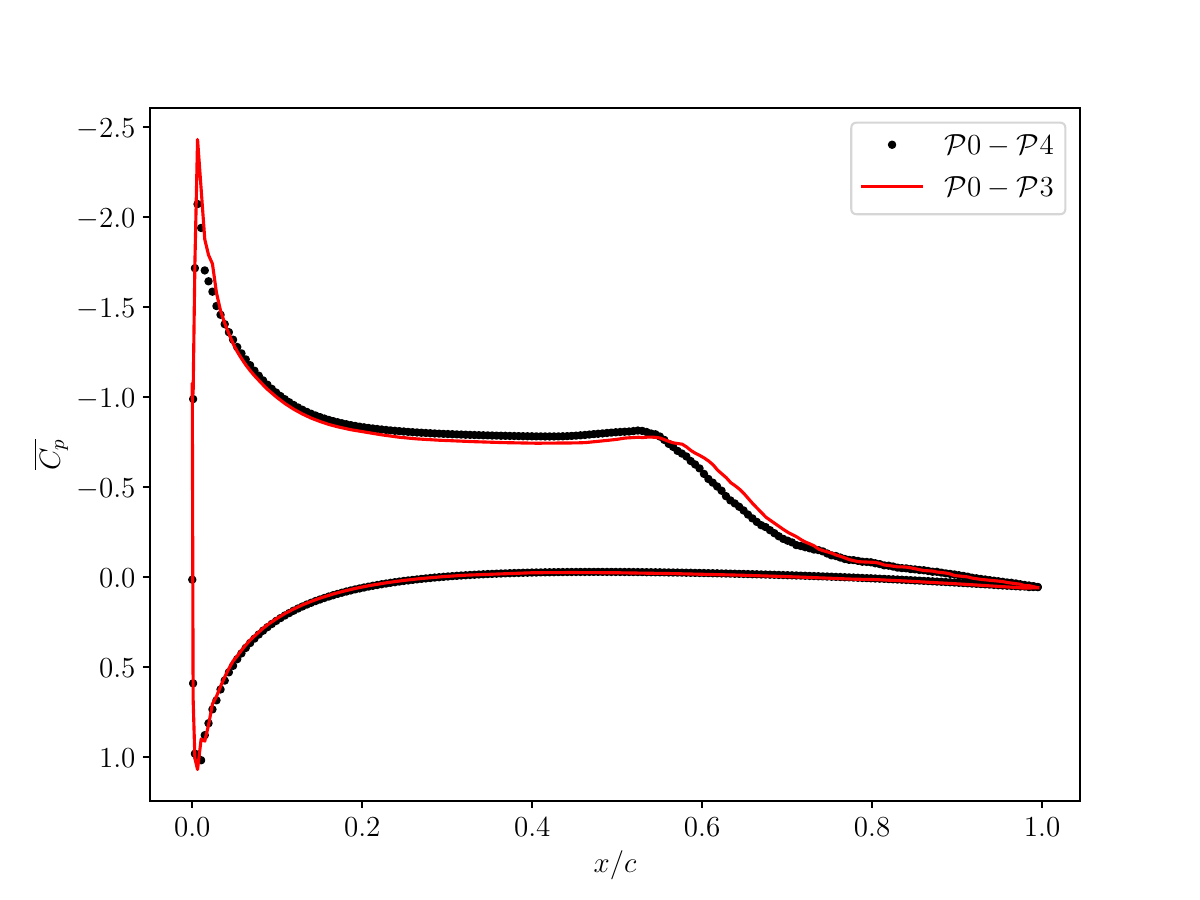}
\caption{The time-averaged pressure coefficient for both $\mathcal{P}3$ and $\mathcal{P}4$ simulations.}
\label{fig_naca_cp}
\end{figure}

\begin{figure}
\centering
\includegraphics[width=0.7\textwidth]{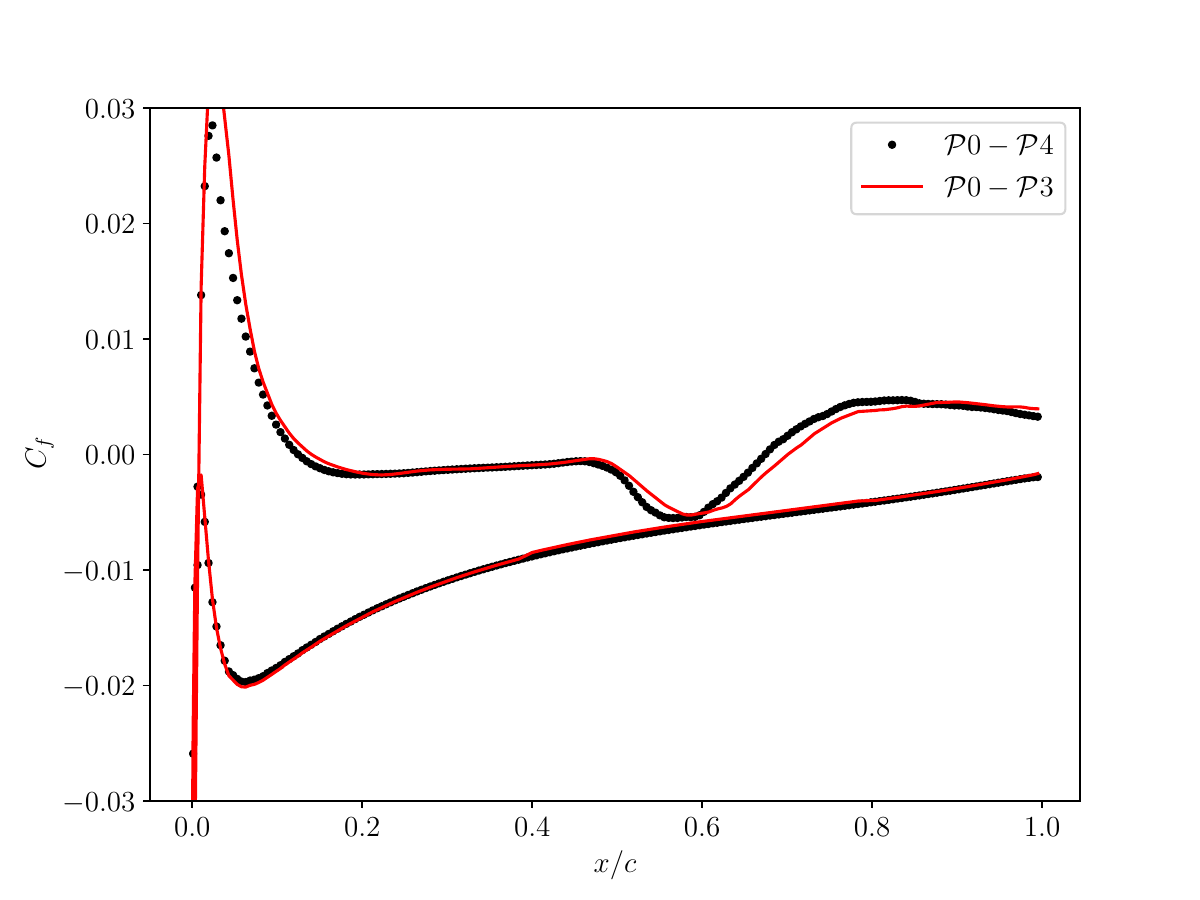}
\caption{The skin friction coefficient for both $\mathcal{P}3$ and $\mathcal{P}4$ simulations.}
\label{fig_naca_cf}
\end{figure}

Two distinct grid resolutions are employed with maximum solution polynomial degrees of $\mathcal{P}3$ and $\mathcal{P}4$. The time-averaged lift and drag coefficients are compared to the ILES reference data \cite{kojima2013large}, presented in Table \ref{table_naca_3d_timehistories}. The difference between the time-averaged lift coefficient obtained from the $\mathcal{P}4$ simulation and the reference data is minimal, affirming the adequacy of the $\mathcal{P}4$ simulation's grid resolution. Furthermore, the time-averaged drag coefficient differs by less than $1.3\%$ from the reference data. The OASPL at an observer located two unit chord lengths below the trailing edge is computed for both $\mathcal{P}3$ and $\mathcal{P}4$ simulations. Various averaging window lengths are applied, and the results are summarized in Table \ref{table_naca_3d_spl_independence}. It is evident that the $\mathcal{P}$ and $\mathcal{P}3$ simulations differ by only $0.5~dB$. The time-averaged pressure coefficient, $\overline{C_p}$, and the skin friction coefficient, $C_f$, for both resolutions are shown in Figures \ref{fig_naca_cp} and \ref{fig_naca_cf}, respectively. These plots show that the separation point, identified with each simulation, are very close and differ by less than $2\%$. Considering the findings presented in Tables \ref{table_naca_3d_timehistories} and \ref{table_naca_3d_spl_independence}, and Figures \ref{fig_naca_cp} and \ref{fig_naca_cf}, we opt to conduct $\mathcal{P}4$ simulation for a total duration of $70$ convective times for the optimization study.

\begin{table}
\centering
\caption{The time-averaged lift and drag coefficients of NACA0012 airfoil at $\alpha = 6^\circ$.}
\begin{tabular}{cccc}
\hline
 & $\mathcal{P}0 - \mathcal{P}3$ & $\mathcal{P}0 - \mathcal{P}4$ & reference \cite{kojima2013large} \\
\hline
$\overline{C_L}$ & $0.6534$ & $0.6399$ & $0.6402$ \\
$\overline{C_D}$ & $0.0553$ & $0.0548$ & $0.0541$ \\
\hline
\end{tabular}
\label{table_naca_3d_timehistories}
\end{table}

\begin{table}
\centering
\caption{The grid independence study of OASPL using different averaging window lengths for NACA0012 airfoil at $\alpha = 6^\circ$.}
\begin{tabular}{ccc}
\hline
\multirow{2}{*}{Averaging Window Length} & \multicolumn{2}{c}{OASPL in $dB$} \\
 & $\mathcal{P}0 - \mathcal{P}3$ & $\mathcal{P}0 - \mathcal{P}4$ \\
\hline
$20t_c$ & $114.9$ & $116.3$ \\
$40t_c$ & $115.7$ & $116.3$ \\
$60t_c$ & $115.7$ & $116.2$ \\
$80t_c$ & $115.7$ & $116.2$ \\
\hline
\end{tabular}
\label{table_naca_3d_spl_independence}
\end{table}

\subsection{Optimization}

The design parameters are maximum camber $c_{max}^{a}$ and its location $x_{c_{max}^{a}}$, maximum thickness $t_{max}^{a}$, and angle of attack $\alpha$, i.e. $\pmb{\mathcal{X}} = [c_{max}^{a}, x_{c_{max}^{a}}, t_{max}^{a}, \alpha]$. The maximum camber range is set to $c_{max}^{a} \in [-10, 10]$ as a percentage of the chord, with the distance from the airfoil leading edge in the range of $x_{c_{max}^{a}} \in [4, 9]$ as a tenth of the chord. The maximum thickness of the airfoil is within the range of $t_{max}^{a} \in [6,18]$ as a percentage of the chord. Finally, the angle of attack varies from $0^\circ$ to $12^\circ$. The objective function is defined as the OASPL at the observer with constraints on both the mean lift and mean drag coefficients. A quadratic penalty term is added to the objective function when the lift coefficient deviates from the baseline design, and an additional quadratic penalty term is added when the mean drag coefficient is above the baseline design. The objective function is defined as
\begin{align}
&
\mathcal{F} = 
\begin{cases}
\text{OASPL} + \epsilon_1 \left( \overline{C_L} - \overline{C_{L,baseline}} \right)^2 + \epsilon_2 \left( \overline{C_D} - \overline{C_{D,baseline}} \right)^2 & \overline{C_D} > \overline{C_{D,baseline}} \\
\text{OASPL} + \epsilon_1 \left( \overline{C_L} - \overline{C_{L,baseline}} \right)^2 & \overline{C_D} \leq \overline{C_{D,baseline}}  \\
\end{cases}
&
,
\label{eq_obj_naca_3d}
\end{align}
where the constants $\epsilon_1$ and $\epsilon_2$ are set to $8,000$ and $400,000$, respectively, to compensate for the order of magnitude difference in OASPL and $\overline{C_L}$ and $\overline{C_D}$. The defined objective function minimizes the OASPL while maintaining the mean lift coefficient, and ensures the optimized airfoil has a similar or lower mean drag coefficient.

\subsubsection{Results and Discussion}

\begin{figure}
\centering
\includegraphics[width=0.5\textwidth]{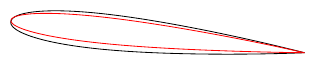}
\caption{The baseline, in black, and optimum, in red, designs of the NACA 4-digits airfoil.}
\label{fig_baseline_optimum_naca_3d}
\end{figure}

\begin{figure}
\centering
\begin{subfigure}{0.75\textwidth}
\centering
\includegraphics[width=\textwidth]{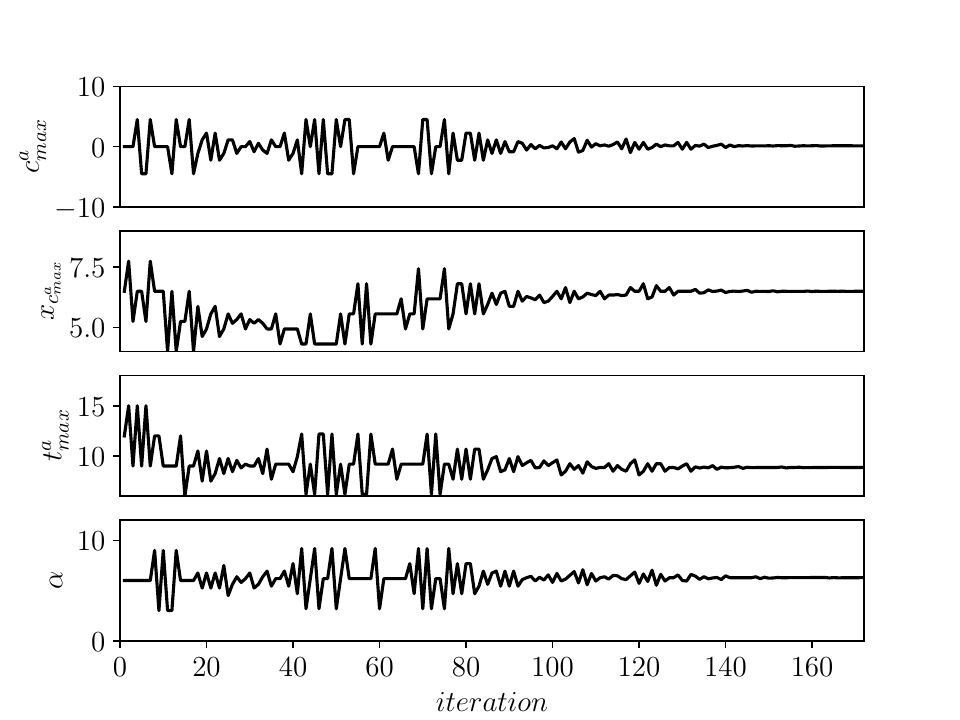}
\subcaption{The design space.}
\end{subfigure}
\begin{subfigure}{0.75\textwidth}
\centering
\includegraphics[width=\textwidth]{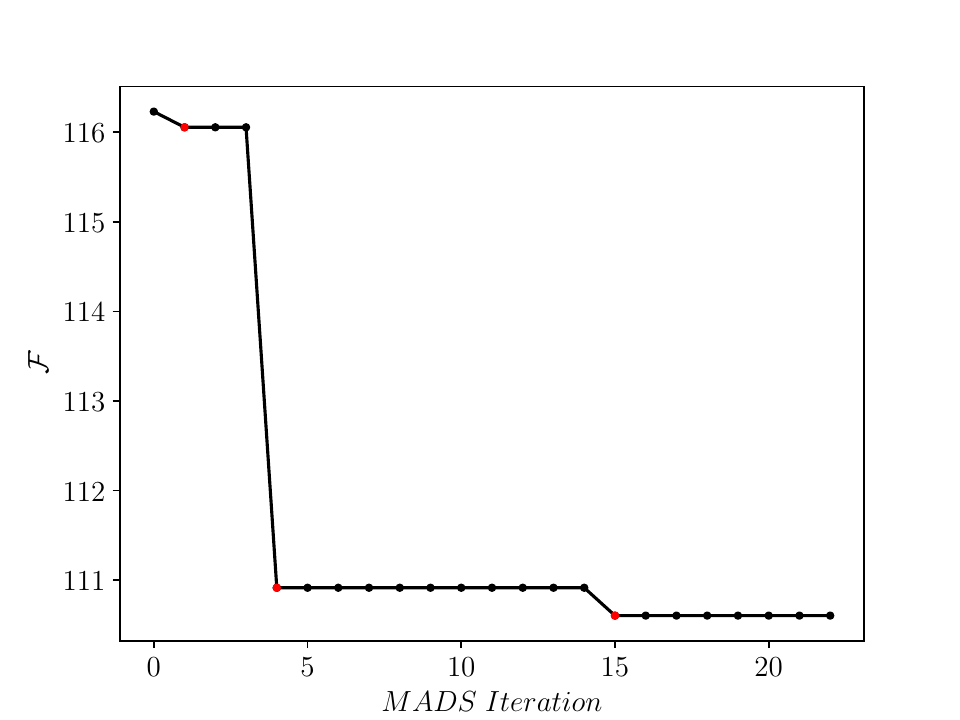}
\subcaption{The objective function convergence with the new incumbent design highlighted in red.}
\end{subfigure}
\caption{The design space and objective function convergence of the NACA 4-digit airfoil optimization.}
\label{fig_naca_3d_optimization_convergence}
\end{figure}

This optimization procedure converges after $22$ MADS iterations, consisting of $172$ objective function evaluations. The baseline and optimum designs, depicted in Figure \ref{fig_baseline_optimum_naca_3d}, demonstrate that reducing the airfoil thickness correlates with lower noise emission. This aligns with the expectation that moving less air leads to reduced noise levels. However, decreasing the airfoil's thickness raises concerns about structural integrity and increases the risk of flutter. The design space and the convergence of the objective function are shown in Figure \ref{fig_naca_3d_optimization_convergence}. The optimal airfoil design has a maximum camber of $c_{max}^{a} = 0.140625$ percent of the chord, at a $6.5$ tenth of the chord distance from the leading edge, with a thickness of $t_{max}^{a} = 8.859375$ percent of the chord, at an angle of attack of $\alpha = 6.28125$ degrees. The OASPL of the optimized airfoil is decreased to $110.6~dB$, the mean lift coefficient is $\overline{C_L} = 0.6556$, and finally, the mean drag coefficient is decreased by $7.4\%$ to $\overline{C_D} = 0.0509$.

\begin{figure}
\centering
\begin{subfigure}{\textwidth}
\includegraphics[width=\textwidth]{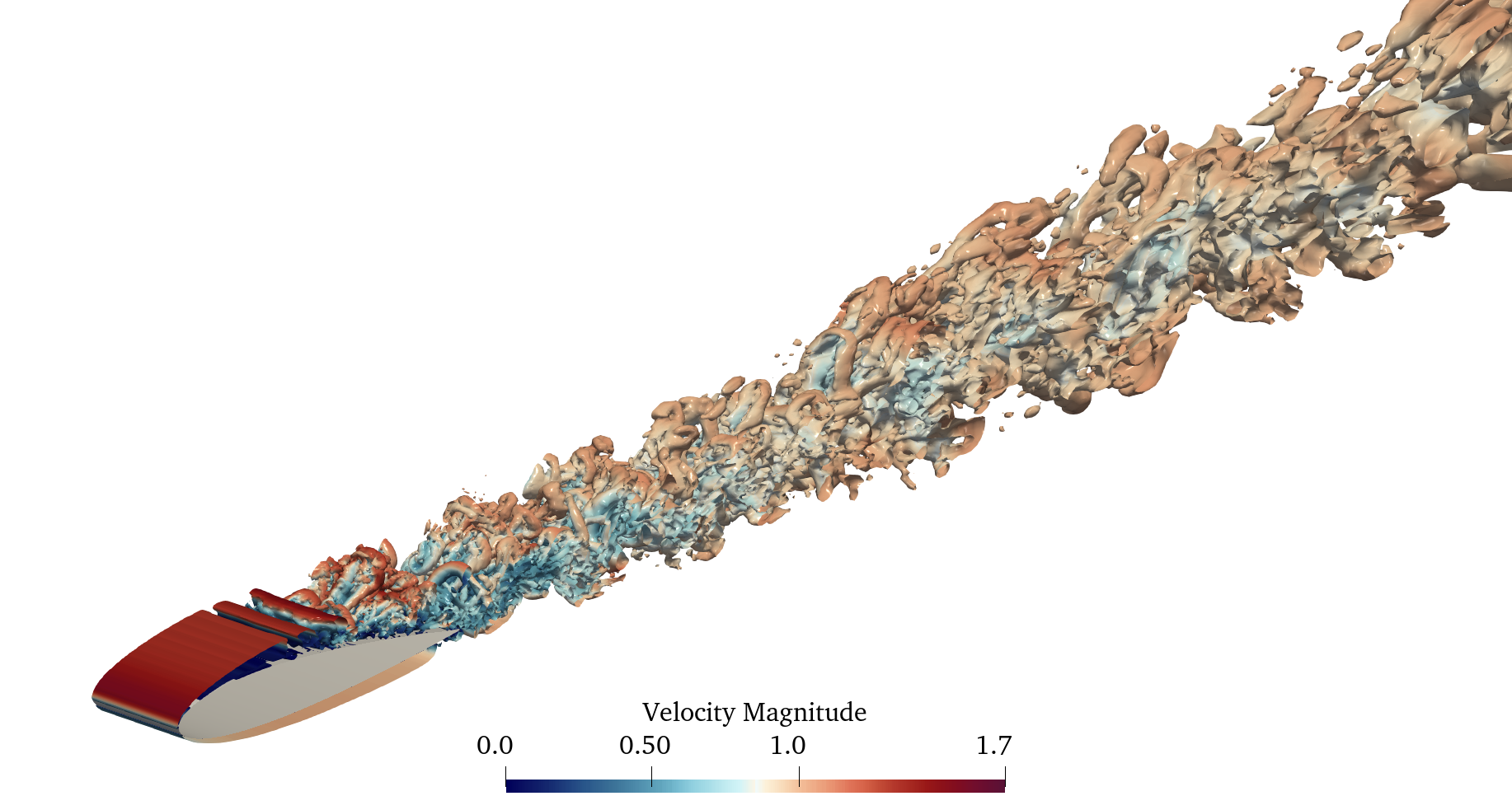}
\subcaption{Q-criterion coloured by velocity magnitude.}
\end{subfigure}
\begin{subfigure}{\textwidth}
\includegraphics[width=\textwidth]{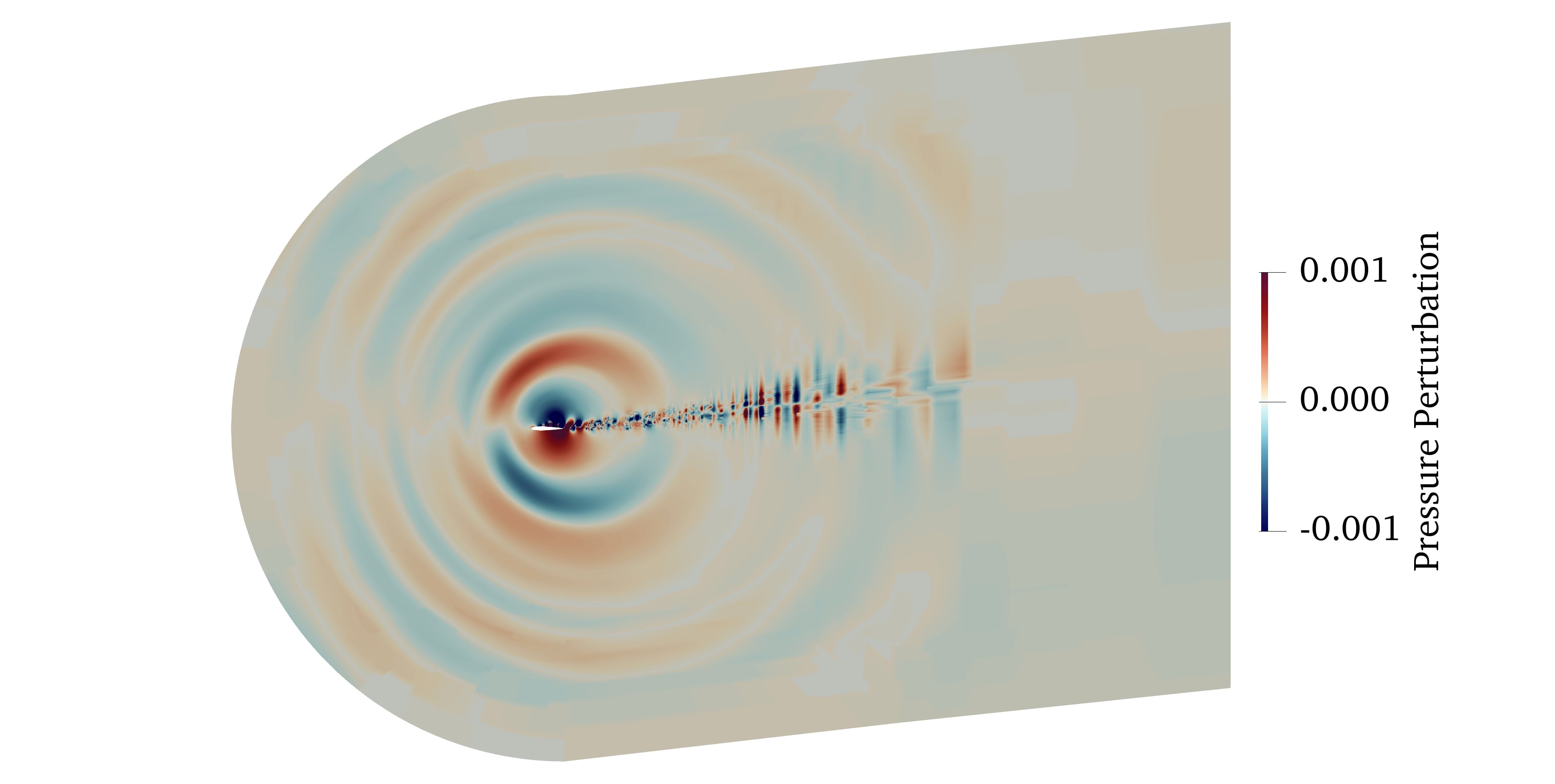}
\subcaption{Acoustic pressure field at mid-plane.}
\end{subfigure}
\caption{The baseline airfoil at $t_c=70$.}
\label{fig_naca_3d_optimization_initial_design}
\end{figure}

\begin{figure}
\centering
\begin{subfigure}{\textwidth}
\includegraphics[width=\textwidth]{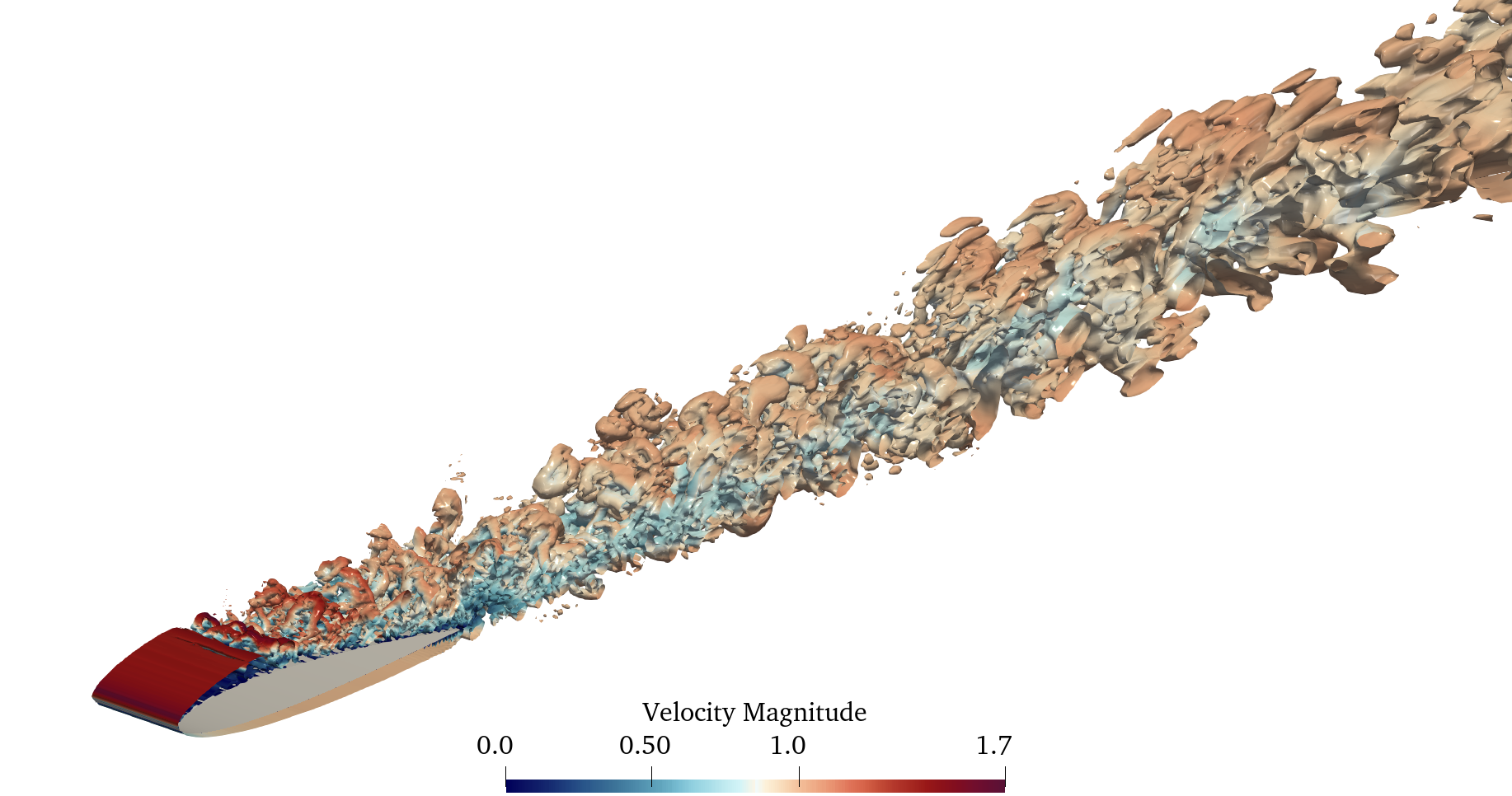}
\subcaption{Q-criterion coloured by velocity magnitude.}
\end{subfigure}
\begin{subfigure}{\textwidth}
\includegraphics[width=\textwidth]{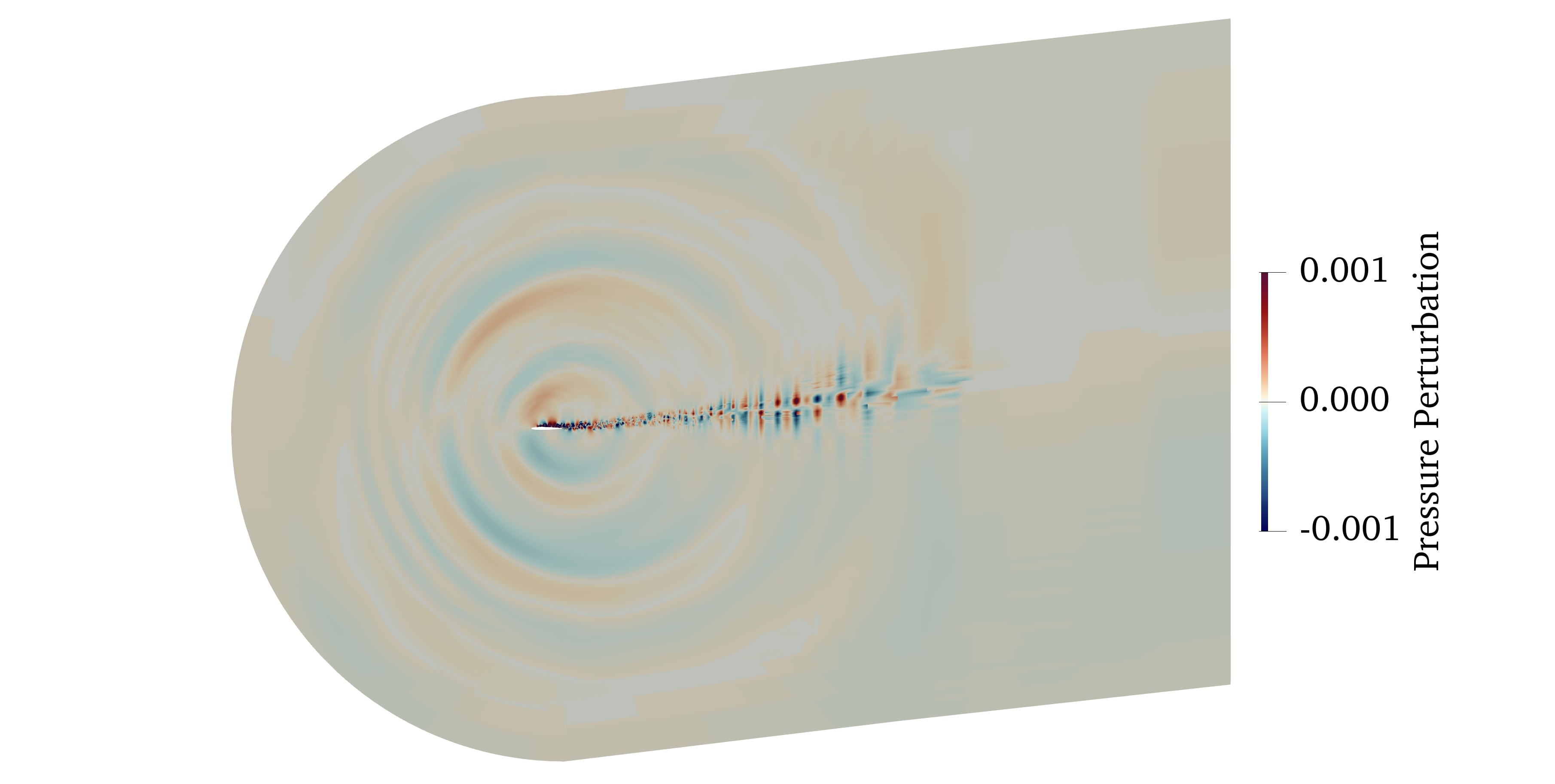}
\subcaption{Acoustic pressure field at mid-plane.}
\end{subfigure}
\caption{The optimum airfoil at $t_c=70$.}
\label{fig_naca_3d_optimization_optimum_design}
\end{figure}

\begin{figure}
\centering
\includegraphics[width=0.7\textwidth]{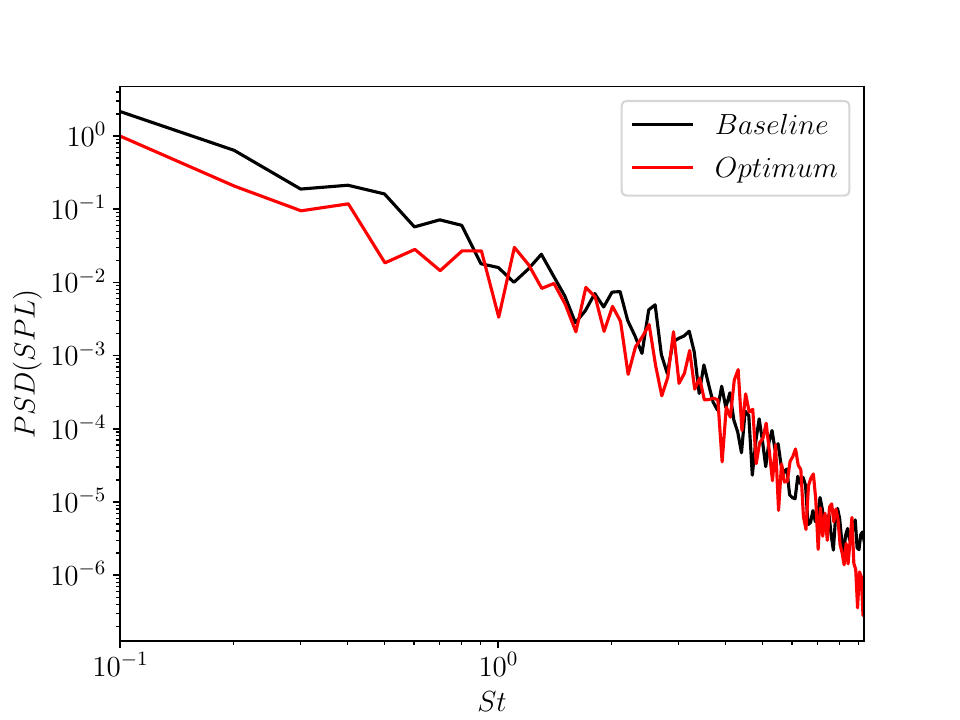}
\caption{The sound spectra for the NACA $4$-digit airfoils.}
\label{fig_naca_3d_sound_spectra}
\end{figure}

Figures \ref{fig_naca_3d_optimization_initial_design} and \ref{fig_naca_3d_optimization_optimum_design} depict the Q-criterion colored by velocity magnitude and pressure perturbation at mid-planes for the baseline and optimum designs. In the optimum design, the separation point shifts towards the leading edge, yielding smaller and less energetic structures, resulting in reduced noise emission. This leads to a $5.7~dB$ decrease in OASPL at a near-field observer. Figure \ref{fig_naca_3d_sound_spectra} presents the PSD of OASPL as a function of the Strouhal number, computed using Welch's method of periodograms \cite{welch1967use} with $3$ windows and a $50\%$ overlap. It is evident that the optimum design displays lower-intensity OASPL energy across various frequency ranges.

\section{Conclusions}
\label{sec:Conclusions}

In conclusion, we present an aeroacoustic shape optimization framework using the MADS optimization algorithm in conjunction with high-order FR spatial discretization and LES. Our framework effectively reduces OASPL at a near-field observer. A key contribution of this work is the elimination of runtime dependency on the number of design parameters through parallel implementation. This addresses a key challenge in gradient-free optimization techniques, enhancing the robustness and computational efficiency of our framework. These findings hold significant importance for aeroacoustic shape optimization, with potential applications in the aerospace industry where noise reduction is of paramount importance. 

It is important to acknowledge that the current study considers a maximum of four design parameters in the airfoil case. While the framework demonstrates consistent runtime for each optimization iteration, equivalent to a single CFD simulation with sufficient computational resources, the scalability of this approach to hundreds or thousands of parameters requires further exploration. It should also be noted that the computational cost of adjoint-based optimization methods is inherently independent of the number of design parameters, making them an attractive option for high-dimensional problems. However, adjoint methods are known to exhibit instability when applied to LES \cite{blonigan2017toward}, limiting their applicability in such contexts. While the proposed framework does not achieve a computational cost that is entirely independent of the number of design parameters, it does facilitate optimization in constant time, assuming adequate parallel resources are available.

The feasibility of the proposed aeroacoustic shape optimization framework can be assessed through testing at higher Reynolds numbers and addressing industry-relevant problems. Additionally, exploring the integration of a far-field acoustic solver into the framework is a promising avenue, potentially broadening its capability to address a more extensive range of aeroacoustic challenges. Furthermore, future work can focus on integrated sound pressure levels across observer surfaces rather than discrete points, aligning the framework more closely with practical applications. This research suggests the potential for more efficient aeroacoustic shape optimization methods, with notable implications for quieter and more efficient aerodynamic designs. Moreover, incorporating design limits for optimization constraints, such as structural failure and flutter phenomenon in airfoil optimization, presents a crucial consideration for enhancing the robustness and applicability of the framework in real-world engineering scenarios.

\clearpage

\section*{Data Statement}

Data relating to the results in this manuscript can be downloaded from the publication’s website under a CC-BY-NC-ND 4.0 license.

\section*{CRediT authorship contribution statement}

\textbf{Mohsen Hamedi:} Conceptualization; Data curation; Formal analysis; Investigation; Methodology; Software; Validation; Visualization; Writing - original draft. 
\textbf{Brian Vermeire:} Conceptualization; Funding acquisition; Investigation; Methodology; Project administration; Resources; Software; Supervision; Writing - review \& editing.

\section*{Declaration of competing interest}

The authors declare that they have no known competing financial interests or personal relationships that could have appeared to influence the work reported in this paper.

\section*{Acknowledgements}

The authors acknowledge support from the Natural Sciences and Engineering Research Council of Canada (NSERC) [RGPIN-2017-06773] and the Fonds de recherche du Québec (FRQNT) via the nouveaux chercheurs program. This research was enabled in part by support provided by Calcul Qu\'ebec (www.calculquebec.ca) and the Digital Research Alliance of Canada (www.alliancecan.ca) via a Resources for Research Groups allocation. M.H acknowledges Fonds de Recherche du Qu\'ebec - Nature et Technologie (FRQNT) via a B2X scholarship.

\bibliography{arXiv}

\end{document}